\documentclass[aps,prd,english,superscriptaddress, longbibliography,11pt,notitlepage]{revtex4}

\usepackage[colorlinks=true, a4paper=true, pdfstartview=FitV,
linkcolor=blue, citecolor=blue, urlcolor=blue]{hyperref}

\usepackage{amsmath}
\usepackage{amssymb}
\usepackage{graphicx}
\usepackage{hhline}
\usepackage{mwe}
\usepackage{amsbsy}
\usepackage{textcomp}
\usepackage{commath}
\usepackage{slashed}
\usepackage{natbib}
\usepackage{subfigure}

\makeatletter
\usepackage[english]{babel}
\numberwithin{equation}{section}
\begin{document}
\immediate\write16{<<WARNING: LINEDRAW macros work with emTeX-dvivers
                    and other drivers supporting emTeX \special's
                    (dviscr, dvihplj, dvidot, dvips, dviwin, etc.) >>}

\title{ On the localized and delocalized modes in kink-antikink interactions: a toy model }
\author{Carlos E. S. Santos}
\email{carlos.santanasantos@ufpe.br}
\affiliation{Departamento de Física, Universidade Federal de Pernambuco, Recife - PE - 50670-901, Brazil}
\author{João G. F. Campos}
\email{joao.gfc@ufma.br}
\affiliation{Física de Materiais, Universidade de Pernambuco, Recife - PE - 50720-001, Brazil}
\author{Azadeh Mohammadi}
\email{azadeh.mohammadi@ufpe.br}
\affiliation{Departamento de Física, Universidade Federal de Pernambuco, Recife - PE - 50670-901, Brazil}
 
\begin{abstract}

This study deals with a piecewise $\phi^2$ scalar field theory in $(1+1)$ dimensions. The scalar field potential is designed with a triple-well shape, engendering kink solutions with asymmetric square-well linearized potentials. Thus, the localized and delocalized modes in this model can be obtained analytically in terms of transcendental equations. This allows us to explore kink-antikink and antikink-kink collisions with any desired number of localized and delocalized modes. We obtain new scenarios of resonance windows suppression, shedding light on the role of higher excited modes in kink scattering.

\end{abstract}

\maketitle

\section{Introduction}
	Topological defects play an important role throughout most, if not all, physical areas. They are intimately connected to the phenomena of phase transition and spontaneous symmetry breaking \cite{doi:10.1142/S0217751X1430018X}. In this picture, they can be understood as the place where the field reaches a symmetric local maximum configuration in order to respect its boundary arrangement. These defects appear in various contexts, including superconducting materials \cite{Yanagisawa:2013aua, YANAGISAWA20183483}, optical phenomena \cite{COULLET1989403, PhysRevLett.119.013902, BRANDAO20164013, Suchkov2016, OZISIK2023170548}, dilaton gravity models \cite{ZHONG2021136716, lima2023aspects}, early universe inflationary models \cite{KIBBLE1980183}, electronic structure in cis/trans polyacetylene \cite{PhysRevB.40.6285, Bernasconi2015}, and magnetic skyrmions, which are topological solitons in chiral magnetic materials \cite{PhysRevB.102.144422, Kuchkin2023, Livramento2023}. In liquid crystals, for instance, the topological defect manifests as disclinations in the nematic phase \cite{kamenskii1985sov, hu2006nonlocality, pu2013rigid, panayotaros2014solitary}.
	
Perhaps the simplest topological defects are kinks, solutions of $(1+1)$ dimensional field theories. Interactions between kinks have been extensively studied since the pioneering works of Campbell et al. \cite{10.1143/PTP.61.1550, Campbell:1983xu, CAMPBELL1986165}. These interactions can result in various outcomes. Most commonly, a kink-antikink pair either reflects to infinity or annihilates, forming a slowly decaying bion. However, in many cases, the interaction can lead to a transient bound state where the pair repeatedly collides before eventually reflecting off to infinity. These brief bound states happen for intervals of initial velocity called resonance windows.
The well-known resonant energy exchange mechanism between translational and vibrational modes explains this phenomenon. It was described for the first time in Ref.~\cite{Campbell:1983xu}. It states that the energy exchange occurs between the kinetic and oscillational bound modes of each kink involved in the process. In the last decades, spectral phenomena, such as resonance windows, have been investigated in a wide range of models. One can mention sine-gordon-like models \cite{CAMPBELL1986165, Belendryasova_2019, PhysRevE.60.3305, Dorey2021, CARRETEROGONZALEZ2022106123, Gani2019}, polynomial models \cite{Gani2021, 0303058, PhysRevD.106.125003, Hose2010Inel, dorey2017boundary, Belendryasova2017, Saxena2019, Khare2022, Bazeia2023, BLINOV2022168739}, models with logarithmic potentials \cite{Khare_2020, BELENDRYASOVA2021136776}, wobbling kink collisions \cite{PhysRevD.103.045003, Campos2021, ALONSOIZQUIERDO2022106183}, and models with impurities \cite{10.1007/3-540-54890-4_193, PhysRevE.95.022202, Lizunova2021}.

Another intriguing spectral phenomenon in kink interactions is the appearance of the so-called spectral walls \cite{PhysRevLett.122.241601, PhysRevD.106.105027}. These walls act as a ``barrier" to the motion of an excited kink, arising when its excitation reaches the threshold between discrete modes and continuum states. Interestingly, the phenomenon also appears in the presence of and mediated by fermions \cite{Campos20232}. Similarly, it was demonstrated in Ref.~\cite{Bazeia20222} that fermions can also mediate the energy exchange mechanism. Recently, a more comprehensive theoretical understanding of spectral phenomena has predominantly been achieved through the collective coordinates approach \cite{PhysRevLett.127.071601}.

Interestingly, many exceptions to Campbell's resonant energy exchange mechanism have been found. Perhaps the most remarkable one is observed in the $\phi^6$ model \cite{Dorey2011Kink}. The authors showed that, although there are no vibrational modes for each defect separately, there are resonance windows due to delocalized bound excitations stored in the internal region between a pair of kinks. Therefore, it will be interesting for the present study to divide the resonance windows into two categories: the ones generated by localized modes and those generated by delocalized modes.

One more exception to the resonance exchange mechanism was obtained in Ref.~\cite{DOREY2018117}. The authors showed that quasinormal modes can mediate resonance windows, but only if the decay rate is low. Then, in Ref.~\cite{Campos2020}, the authors identified the same phenomenon considering a piecewise $\phi^2$ model, which allowed for the analytical treatment of quasinormal modes. Scalar field theories with piecewise potentials can serve as effective laboratories for studying more complex realistic models, offering opportunities for analytical analysis that may not be feasible in realistic scenarios. Examples in the literature include the scattering between signum-Gordon oscillons \cite{Klimas2020} and the scattering of kinks in coreless potentials \cite{Klimas2024, Blaschke2024}. They also appear as the limit of some families of polynomial potentials \cite{Xiang2024}.
	
In the present work, we generalize the model designed in Ref.~\cite{Campos2020} with a piecewise $\phi^2$ toy model to the one with a triple-well shape potential and asymmetric kink solutions. It allows us to control the number of localized and delocalized modes in kink interactions. These modes can be determined by solving transcendental equations derived from the Schrödinger-like stability equation. In this fashion, it is possible to conduct an in-depth analysis of the resonance window phenomenon. 

The remaining sections are organized as follows. Section 2 describes our toy model. In Section 3, we derive the linearized potentials for kinks and antikink-kink pairs and analytically determine the bound energy spectrum for both localized and delocalized modes in terms of transcendental equations. Section 4 presents the analysis of our numerical scattering results. Finally, we present and discuss our concluding remarks in Section 5. We work with natural units throughout the paper.

\section{The model}
	
\begin{figure}
		\centering
		\subfigure[]{
		\includegraphics[width=0.48\textwidth]{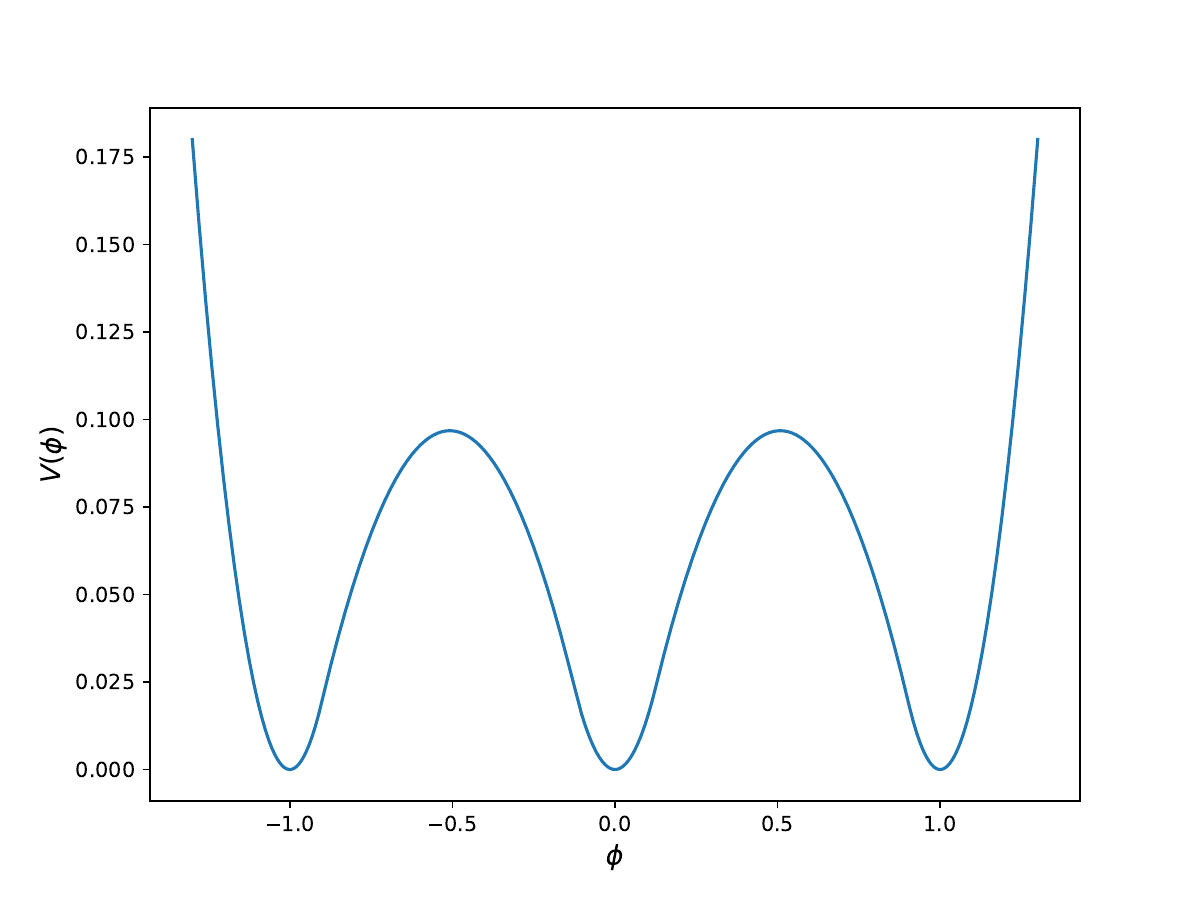}}
		\subfigure[]{
		\includegraphics[width=0.48\textwidth]{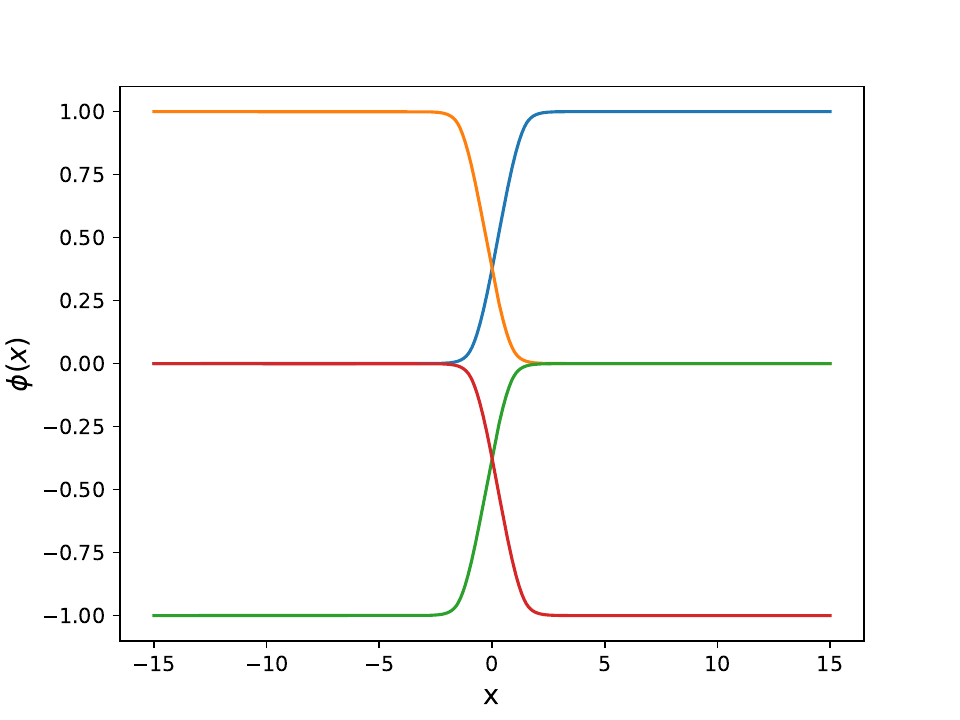}}
		\caption{(a) Piecewise $\phi^2$ potential with three minima. (b) Profiles for all kinks and antikinks of our theory (four topological sectors). Parameters are $A^2 = 3$ and $C^2 = 4$.}
		\label{fig:kink-pot}
	\end{figure}

Let us consider the following scalar field theory in $(1+1)$ dimensions
	\begin{equation}
	\mathcal{L} = \frac{1}{2} \partial_\mu\phi \partial^\mu \phi - V(\phi).
	\end{equation}
Similarly to Ref.~\cite{Campos2020}, the potential $V(\phi)$ is a piecewise $\phi^2$ function. In the current case, it is given by
\begin{equation}
V(\phi) =
\begin{cases}
\frac{A^2}{2} \phi^2, &0 < \phi < \phi_1, \\
- \frac{B^2}{2} (\phi - \phi_0)^2 + V_+, &\phi_1 < \phi < \phi_2, \\
\frac{C^2}{2} (\phi - \lambda)^2, &\phi > \phi_2.
\end{cases}
\end{equation}
The potential for negative $\phi$ is defined by enforcing even parity. Well-behaved kinks are obtained by ensuring the continuity of the potential and its derivative at the boundaries of each domain. The resulting potential, shown in Fig.~\ref{fig:kink-pot}, consists of a triple-well structure similar to the $\phi^6$ model \cite{Dorey2011Kink}.  
There are eight free parameters in this theory, $A$, $B$, $C$, $\phi_0$, $V_+$, $\lambda$, $\phi_1$ and $\phi_2$. The continuity conditions lead to four relations listed in Appendix~\ref{ap:rel}. Thus, only four parameters are independent, which we choose to be $A$, $B$, $C$, and $\phi_0$. Then, we work with dimensionless variables by virtue of the following rescaling $x^\mu\to B^{-1}\phi_0^{-1}x^\mu$ and $\phi\to\phi_0\phi$. Redefining the constants sets $B=1$ and $\phi_0=1$ effectively. Therefore, $A$ and $C$ are the only remaining free parameters.  

The Euler-Lagrange equation for the theory is
	\begin{equation}
	\partial_{tt}\phi - \partial_{xx}\phi = - \frac{dV}{d\phi}.
	\end{equation}
For static field configurations, one obtains the first-order Bogomol'nyi–Prasad–Sommerfield (BPS) equation as follows
	\begin{equation}
	\frac{d \phi_{\text{K}}(x)}{dx} = \pm\sqrt{2 V(\phi_K)}.
	\end{equation}
There are four topological sectors, two kinks and two antikinks, in the theory as shown in Fig.~\ref{fig:kink-pot}. We show the kink sectors by $(-1,0)$, $(0,1)$ and the antikinks by $(0,-1)$, $(1,0)$.  Then, we arrive at the following kink profile
	\begin{equation}
	\label{eq:kink-prof}
	\phi_K(x) =
\begin{cases}
e^{Ax}, &x < x_1,\\
\phi_0 + K \sin [B(x-x_1) + \theta_0], &x_1 < x < x_2,\\
\lambda + (\phi_2 - \lambda) e^{-C(x-x_2)}, &x > x_2.
\end{cases}
	\end{equation}
in $(0,1)$ sector. There are four extra parameters in eq.~\ref{eq:kink-prof}, $K$, $\theta_0$, $x_1$ and $x_2$. We fix them by requiring continuity of the kink function and its derivative at the points $x = x_1$ and $x = x_2$. The relations are listed in Appendix~\ref{ap:rel}. Thus, only $A$ and $C$ remain as free parameters.

\section{Stability equation}

\begin{figure}
		\centering
		\subfigure[]{
		\includegraphics[width=0.48\textwidth]{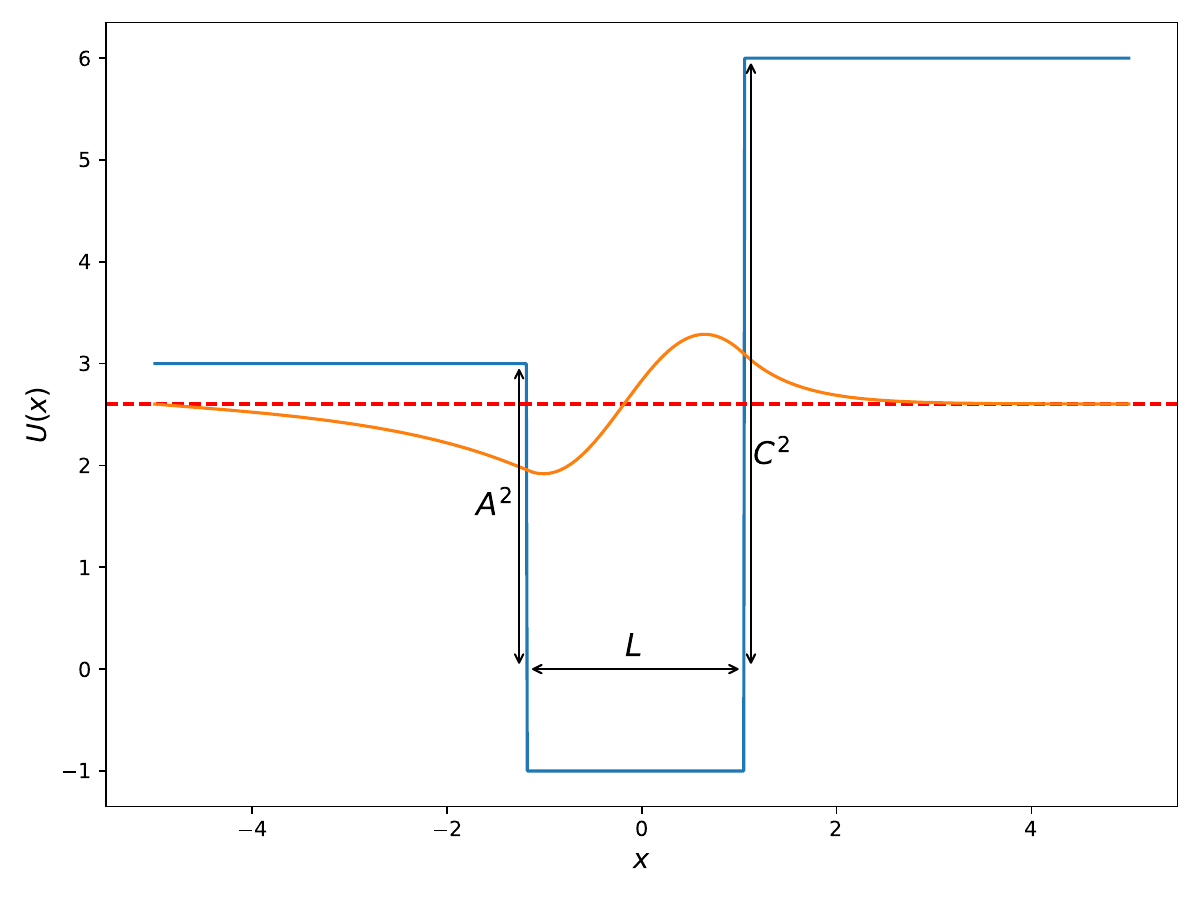}}
		\subfigure[]{
		\includegraphics[width=0.48\textwidth]{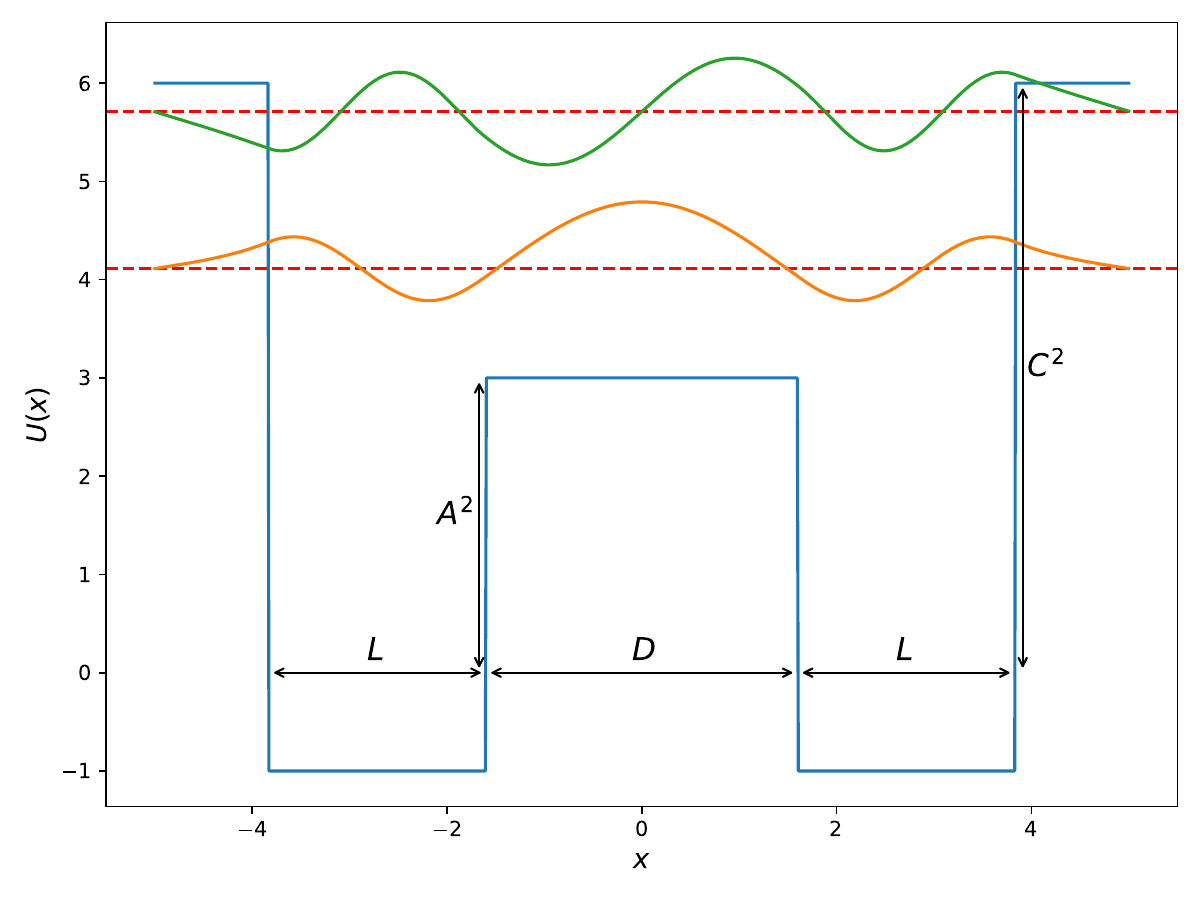}}
		\caption{(a) Linearized potential for a single kink. (b) Linearized potential for antikink-kink pairs. Parameters are $A^2 = 3$, $C^2 = 6$.}
		\label{fig:lin-pot}
	\end{figure}

	The stability of the kinks in our model can be obtained by studying perturbation of the form $\phi(x,t) = \phi_{\text{K}}(x) + \eta(x) e^{-i \omega t}$. Then, we arrive at the following Schrödinger-like equation
	\begin{equation}
	\left[ - \frac{d^2}{dx^2} + U(x) \right] \eta (x) = \omega^2 \eta(x).
	\end{equation}
The linearized potential $U(x)$ is given by
	\begin{equation} \label{stab-pot}
	U(x) = \frac{d^2 V(\phi)}{d\phi^2}\bigg|_{\phi = \phi_K(x)}=\begin{cases}
A^2, &x < x_1, \\
-1, &x_1 < x < x_2, \\
C^2, &x > x_2.
\end{cases}
	\end{equation}
We have constructed a potential with only quadratic functions in order to obtain the asymmetric square-well stability potential, as depicted in Fig.~\ref{fig:lin-pot}(a).
	
\begin{figure}
		\centering
		\includegraphics[width=0.55\textwidth]{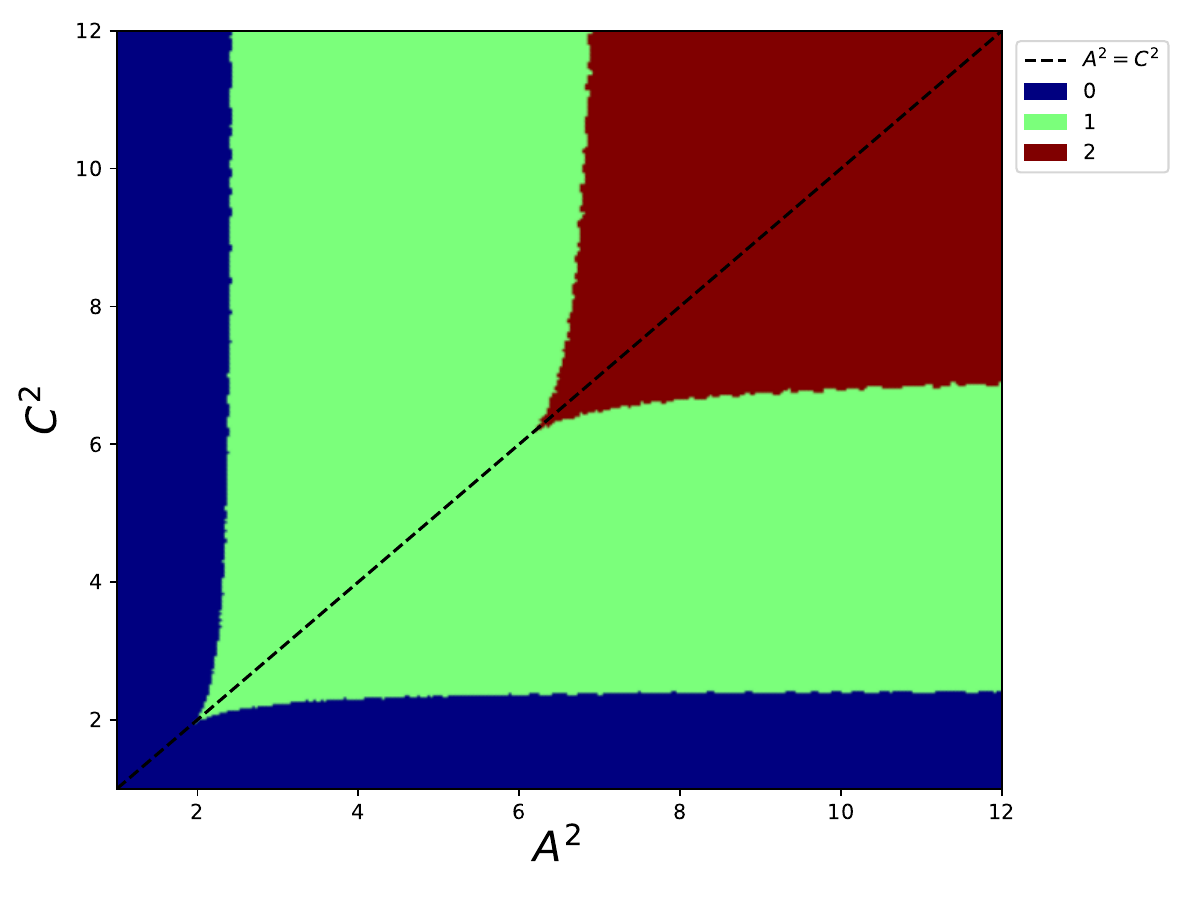}
		\caption{Number of bound modes for an asymmetric square well. The dashed line represents the number of localized modes when $A=C$.}
		\label{fig:modes}
	\end{figure}

We can utilize tools from canonical quantum mechanics to solve the Schrödinger-like stability equation. The linearized potential in eq.~\ref{stab-pot} possesses three types of solutions: localized bound modes with $0\leq\omega^2<A^2$, half-localized modes with $A^2<\omega^2<C^2$ and scattering modes with $\omega^2>C^2$. At this point, we are interested in the localized modes. Thus, we begin by defining the momenta as $k_1^2 = A^2 - \omega^2$, $k_2^2 = 1 + \omega^2$ and $k_3^2 = C^2 - \omega^2$. By applying the continuity conditions for the field and its first derivative, we derive transcendental equations for the energy spectrum. These equations are given by
\begin{equation}\label{trans}
	\tan(k_2 L) = \frac{k_2(k_1 + k_3)}{k_2^2 - k_1 k_3}.
	\end{equation}
After solving for $\omega$, the analytical expression for the corresponding bound modes can be found in Appendix~\ref{ap:modes}. In Fig.~\ref{fig:lin-pot}(a), one can also see the profiles of the corresponding bound solutions for the fixed values $A^2 = 3$, $C^2 = 6$, for example. The number of vibrational shape modes, obtained by solving the above transcendental equation, eq.~\ref{trans}, numerically, is shown in Fig.~\ref{fig:modes} as a function of the free parameters $A$ and $C$. As expected, it increases with both parameters.

It is interesting to take the limit that $C\to\infty$, as a consistency check. In such a case, the linearized potential becomes a semi-infinite square well. Its eigenvalues are well-known, being described by the following transcendental equation
\begin{equation}
\tan(k_2L)=-\frac{k_2}{k_1},
\end{equation}
which agrees with our previous calculation if we set $k_3\to\infty$.

We are also interested in delocalized modes, which are modes that exist in the linearized potential generated by a kink-antikink (or antikink-kink) pair. In our toy model, the profile consists of two asymmetric square wells, as shown in Fig.~\ref{fig:lin-pot}(b). In such a scenario, the delocalized modes can be obtained analytically. They are defined as bound solutions with $A^2<\omega^2<C^2$. If $A^2 > C^2$, the inner barrier becomes higher than the outer ones, and no delocalized modes can exist.

Again, considering the continuity condition and redefining $k_1^2=\omega^2-A^2$, one arrives at the following transcendental equations
	\begin{align} \label{trans-delocalized}
	\tan\left(\frac{k_1D}{2}\right) &= \frac{\frac{k_3}{k_1} -\frac{k_2}{k_1} \tan(k_2 L)}{1 + \frac{k_3}{k_2} \tan(k_2 L)}, \nonumber\\
	\tan\left(\frac{k_1D}{2}\right) &= \frac{\frac{k_1}{k_2} \left[1+\frac{k_3}{k_2} \tan(k_2 L) \right]}{\tan(k_2 L)-\frac{k_3}{k_2}},
	\end{align}
for even and odd delocalized modes, respectively. Once the frequency $\omega$ is computed, the analytical expression of the corresponding bound mode can be obtained as described in Appendix~\ref{ap:modes}. In Fig.~\ref{fig:lin-pot}(b), the delocalized modes of the presented potential for the fixed values $A^2 = 3$, $C^2 = 6$ are also shown. Moreover, the values of $\omega^2$ as a function of $D/2$ for several values of $A^2$ and $C^2$ are shown in Figs.~\ref{fig:spec01}, \ref{fig:spec11} and \ref{fig:spec02}.

Transcendental equations for delocalized modes could also be derived by a novel approach suggested in \cite{long2024solving}, connecting kink and antikinks' half-localized modes, $\eta(x)$, to kink-antikink delocalized modes imposing the conditions
\begin{align} \label{PBA}
\dfrac{\partial \eta (x - D/2)}{\partial x}\bigg|_{x = 0} &= 0,\nonumber\\
\dfrac{\partial^2 \eta (x - D/2)}{\partial x^2}\bigg|_{x = 0} &= 0,
\end{align}
for even and odd modes, respectively. In general, the method gives a good approximation for delocalized modes only when the kink and antikink are separated enough. Interestingly, in our case, applying the conditions (\ref{PBA}) on the half-localized modes given in Appendix~\ref{ap:modes} for $A^2<\omega^2<C^2$ gives the exact results (\ref{trans-delocalized}) for even and odd delocalized modes.

It is also instructive to take the limit $C\to\infty$ in the kink-antikink linearized equation. By doing so, one obtains an infinite square well with a perturbation at its center. If the energy level is high enough, the perturbation becomes negligible. In this limit, we have an infinite square-well plus corrections in powers of $A^2$, yielding a simple form for the delocalized modes frenquencies
\begin{equation}
\omega^2=\frac{n^2\pi^2}{(D+2L)^2}-1+A^2 \left[\frac{D}{D+2L}-\frac{1}{n\pi}\mathrm{sin} \left(\frac{n\pi D}{D+2L}\right)\right]+\mathcal{O}(A^4).
\end{equation}

In the following sections, we are interested in analyzing kink-antikink and antikink-kink collisions for several values of $A^2$ and $C^2$. By probing configurations in this parameter space, we will be able to study scattering scenarios where distinct numbers of localized and delocalized modes are available.
	
\section{Kink-antikink and antikink-kink collisions}
	
Now, let us analyze the kink-antikink (and antikink-kink) collisions for our toy model. The kink-antikink scattering is achieved by solving the field equation with the following initial condition
	\begin{equation}
	\phi(x,0) = \phi_{\text{K}}(\gamma (x + X_0)) + \phi_{\bar{\text{K}}}(\gamma (x - X_0))-1,
	\end{equation}
	\begin{equation}
	\dot{\phi}(x,0) = -\gamma v \left[ \phi^\prime_{\text{K}}(\gamma (x + X_0))  -\phi^\prime_{\bar{\text{K}}}(\gamma (x - X_0)) \right].
	\end{equation}
where $v$ is the initial velocity, $\gamma(v)$ is the Lorentz factor and the subscripts $K$ and $\bar{K}$ refer to the kink and antikink solutions, respectively. On the other hand, the antikink-kink initial conditions are
	\begin{equation}
	\phi(x,0) = \phi_{\bar{\text{K}}}(\gamma (x + X_0)) + \phi_{\text{K}}(\gamma (x - X_0)),
	\end{equation}
	\begin{equation}
		\dot{\phi}(x,0) = -\gamma v \left[\phi^\prime_{\bar{\text{K}}}(\gamma (x + X_0)) - \phi^\prime_{\text{K}}(\gamma (x - X_0)) \right].
	\end{equation}
The details of our numerical algorithm can be found in Appendix~\ref{ap:num}.

To obtain the same number of delocalized modes for kink-antikink and antikink-kink scenarios, we will swap the parameters $A$ and $C$. In other words, we will compute kink-antikink collisions with parameters $(A,C)=(p_1,p_2)$ with $p_1>p_2$. Then, we will repeat the simulations for antikink-kink collisions with $(A,C)=(p_2,p_1)$. To avoid ambiguities, we emphasize that we are referring to kinks and antikinks in $(0,1)$ and $(1,0)$ sectors.

We are interested in whether resonance windows exist in our family of models. Whenever they are present, the frequency of the resonant mode will be obtained by the following relation \cite{Campbell:1983xu}
	\begin{equation}
	\omega T = 2 \pi n + \delta
	\end{equation}
with $T$ being the time between collisions, $n$ the resonance window number and $\delta$ a constant parameter in the interval $[0,2\pi)$. Then, the relative error between the numerical frequency and the analytical solutions of the localized modes is defined as
	\begin{equation}
	\delta \omega =\frac{|\omega_\text{
	theoretical} - \omega_{\text{collision}}|}{\omega_{\text{
	theoretical}}}.
	\end{equation}
It will allow us to identify the resonant localized mode among the spectrum. For delocalized modes, a graphical construction will be presented.

\begin{figure}
    \centering
    \subfigure[ $v=0.15$]{\includegraphics[width=0.36\textwidth]{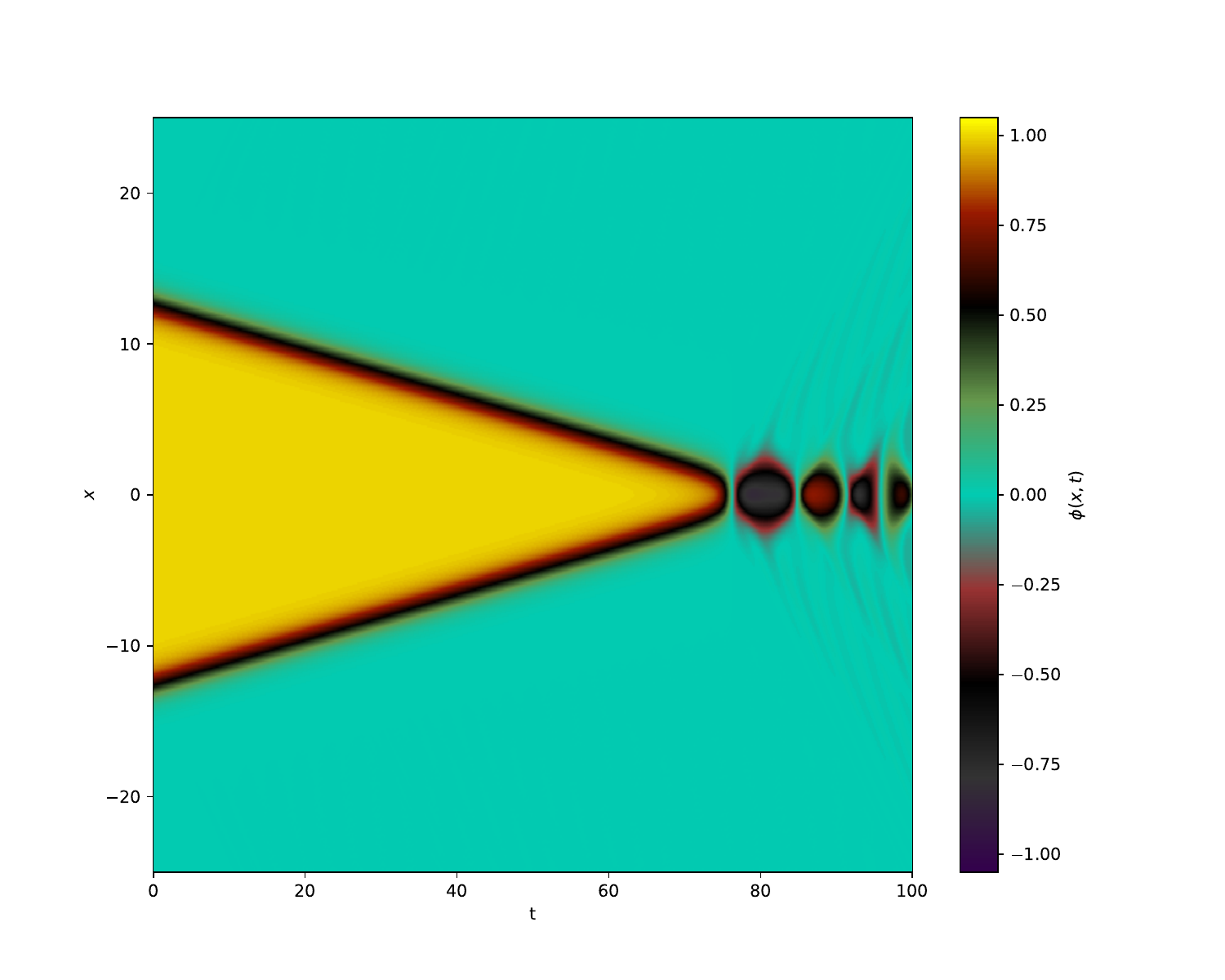}}
    \subfigure[ $v=0.2$]{\includegraphics[width=0.36\textwidth]{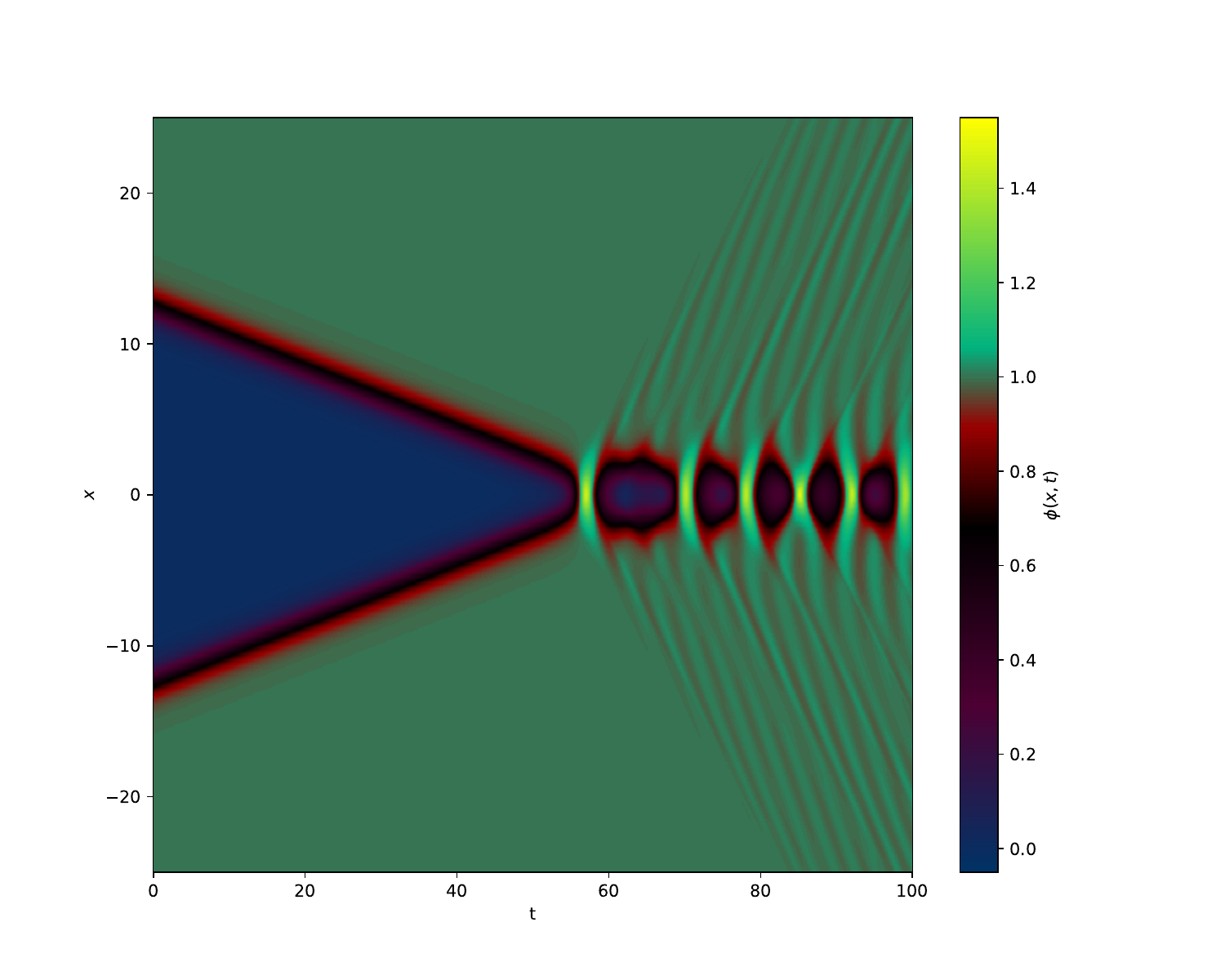}}
    \subfigure[ $v=0.211$]
{\includegraphics[width=0.36\textwidth]{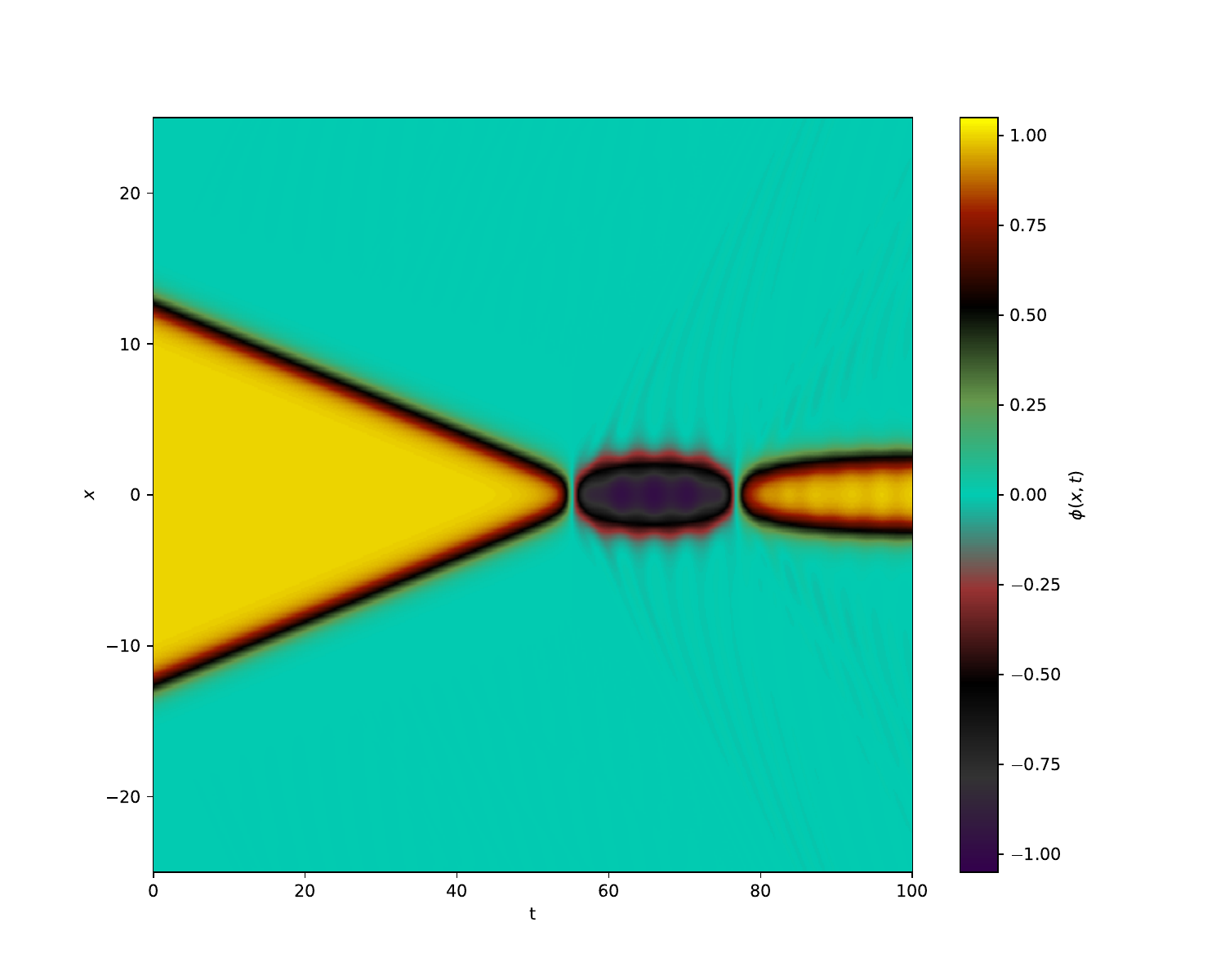}}
    \subfigure[ $v=0.23$]{\includegraphics[width=0.36\textwidth]{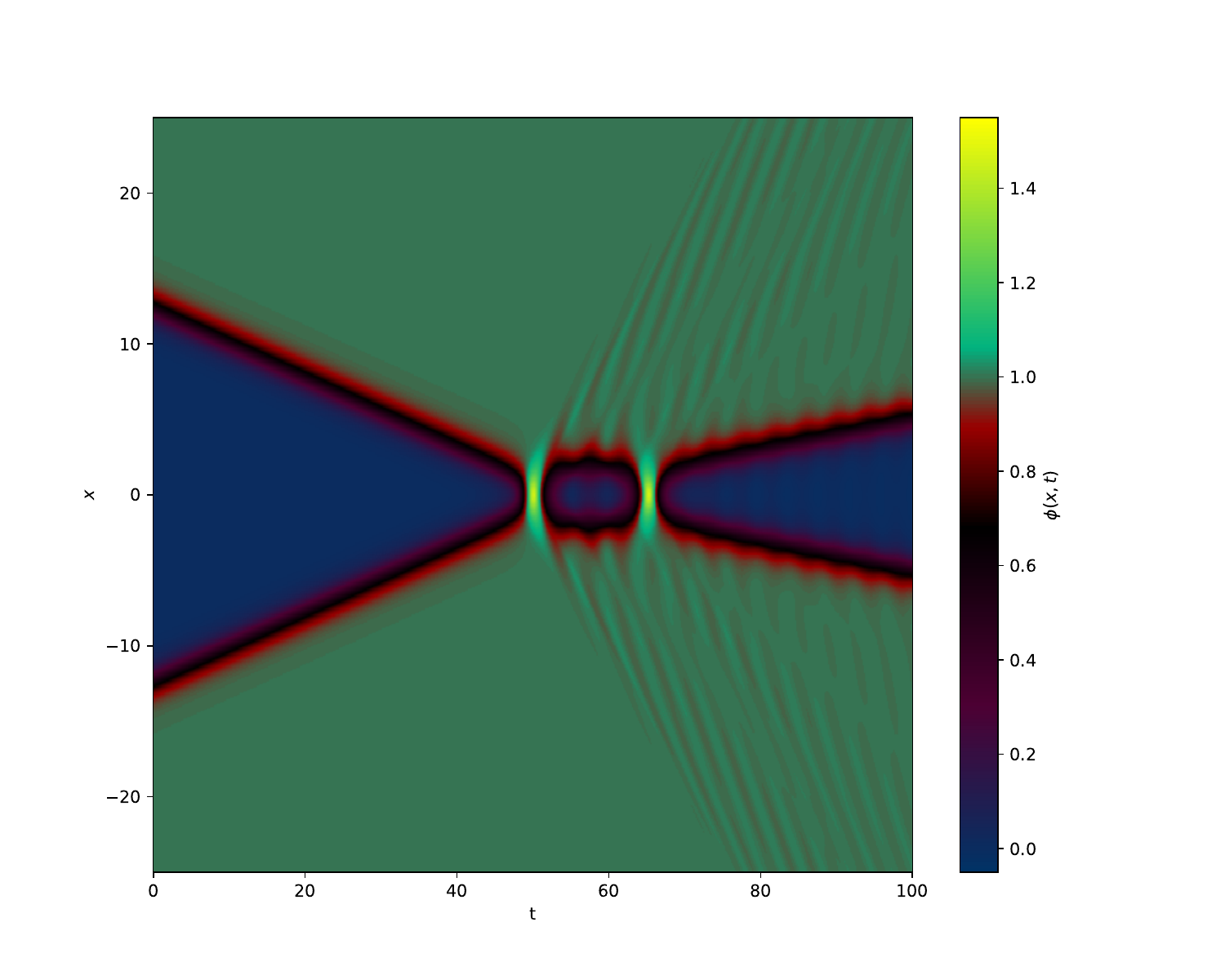}}  
	\subfigure[ $v=0.4$]{\includegraphics[width=0.36\textwidth]{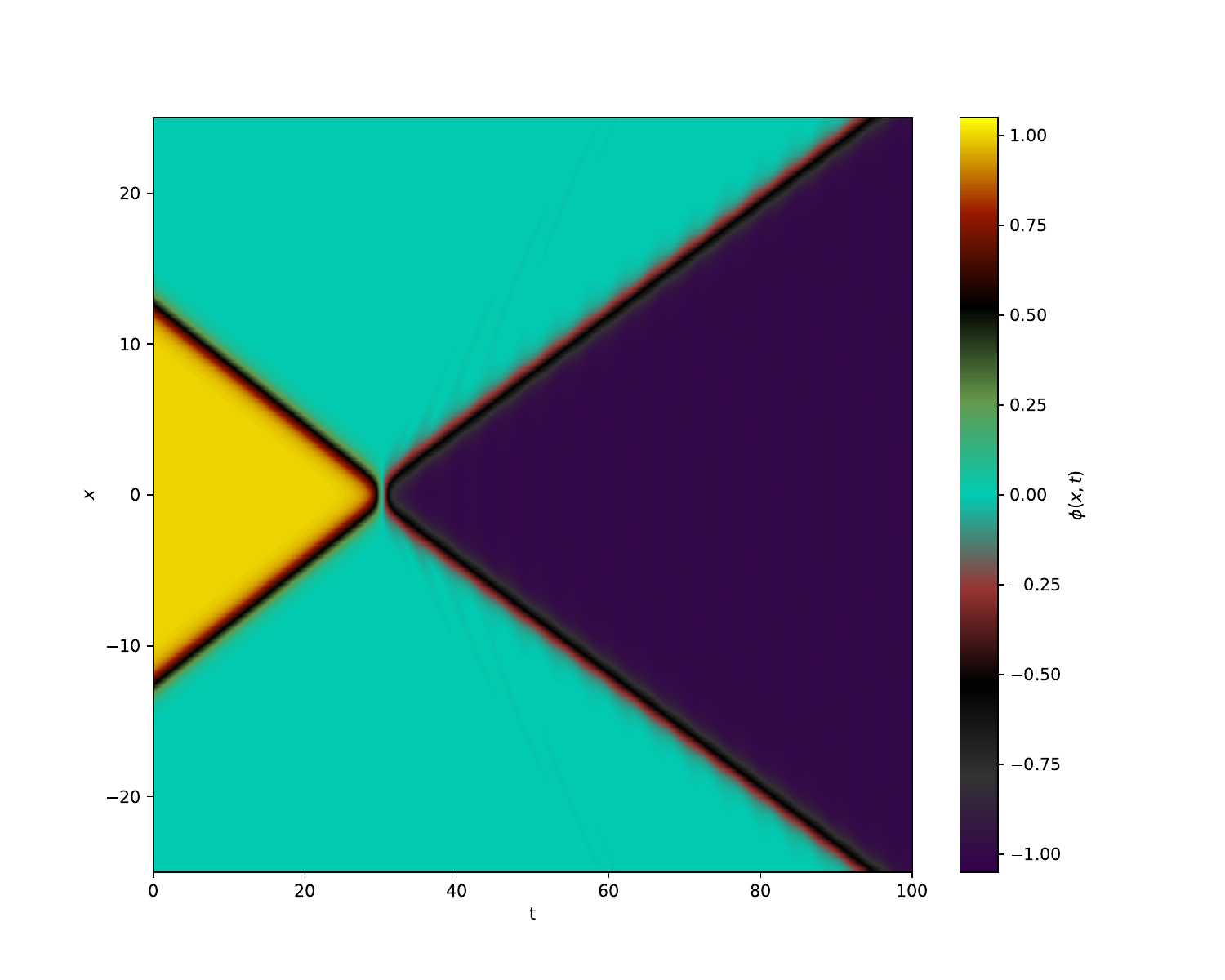}}    
    \subfigure[ $v=0.4$]{\includegraphics[width=0.36\textwidth]{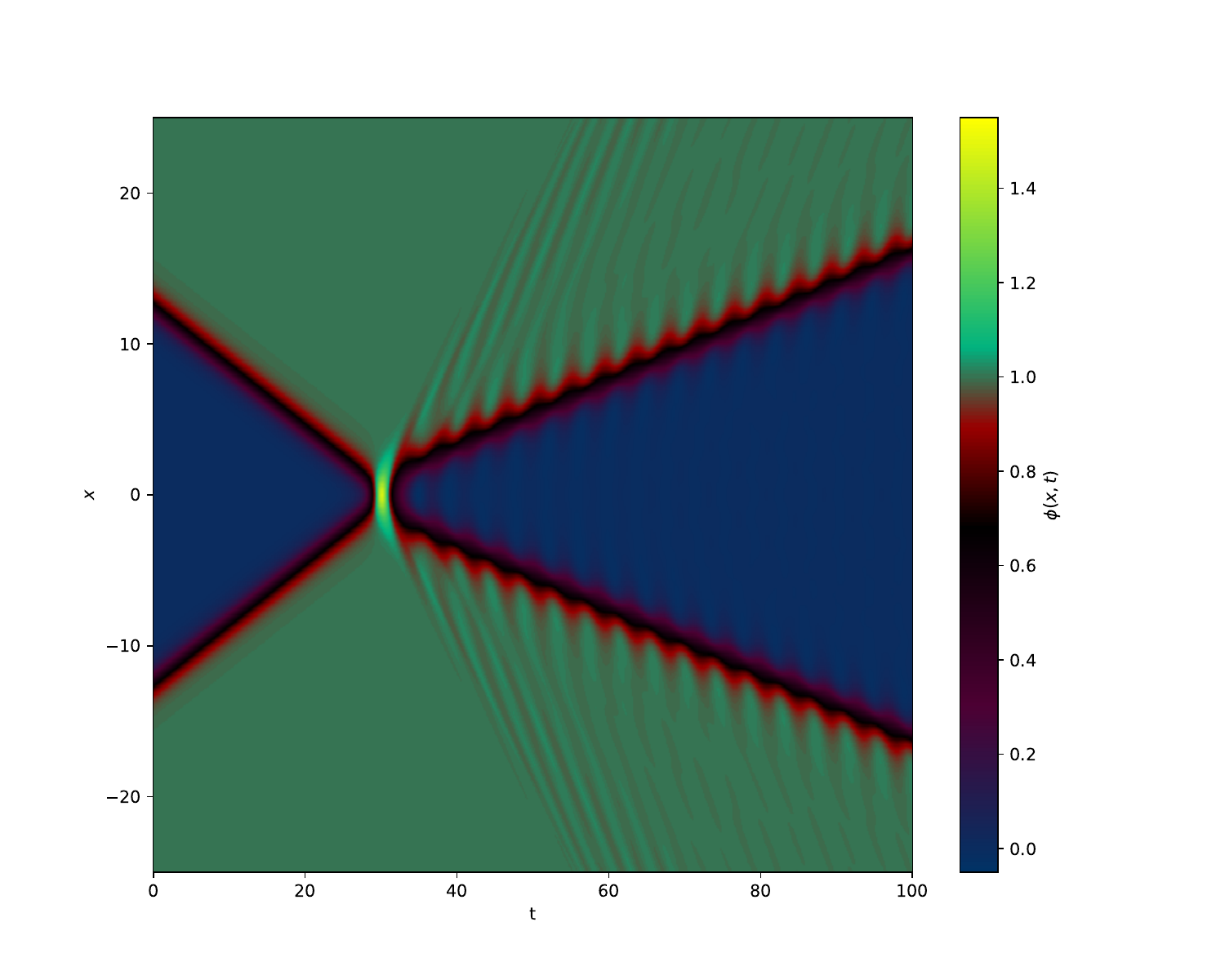}}
    
    \caption{Bion formation in (a) an kink-antikink, and (b) a antikink-kink collision. Two-bounce window for (c) a kink-antikink, and (d) an antikink-kink collision. Reflection for (e) a kink-antikink, and (f) an antikink-kink collision. Parameters are $A^2=C^2=8/3$.}
    \label{fig:cols}
\end{figure}

The allowed scattering outputs are shown in Fig.~\ref{fig:cols}. They consist of bion formation, multi-bounce windows, and reflection. However, unlike antikink-kink collisions, the kink-antikink interaction permits sector change. This means that, at each bounce, the kinks re-emerge in the opposite sector rather than the same one. Notably, in a two-bounce window, the sector change occurs twice. This behavior is characteristic of models with multiple sectors.
	
\subsection{$(A^2, C^2)=(8/3,8/3)$}

The most common resonance scenario is the case where there is one localized mode and no delocalized mode. In our model, this can be achieved by fixing $A^2 = C^2 = 8/3$, for instance. In such a case, the linearized potential in Fig.~\ref{fig:lin-pot}(a) is a symmetric well.
	
\begin{figure}
    \centering  
    \subfigure[]{\includegraphics[width=0.48\textwidth]{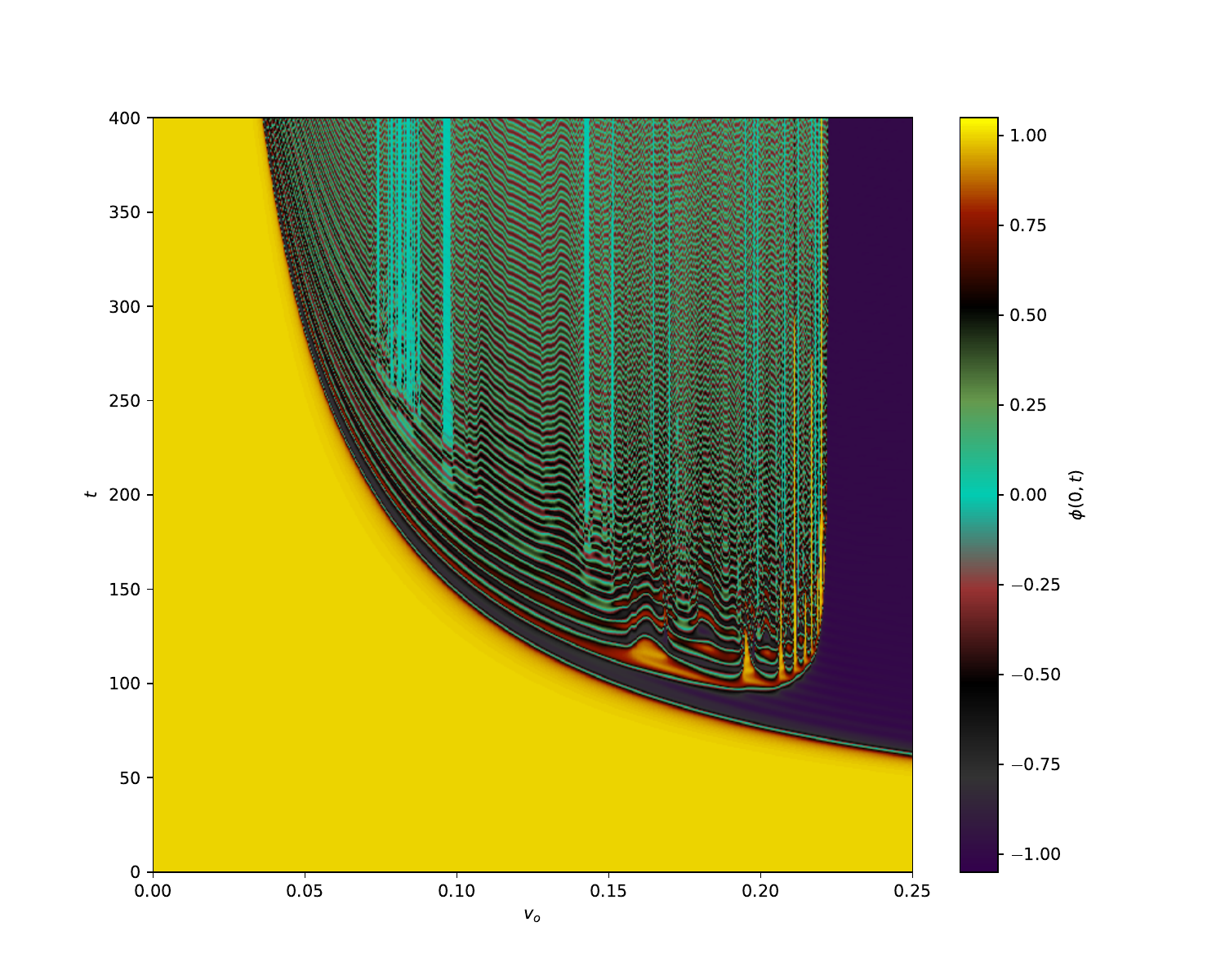}}
    \subfigure[]{\includegraphics[width=0.48\textwidth]{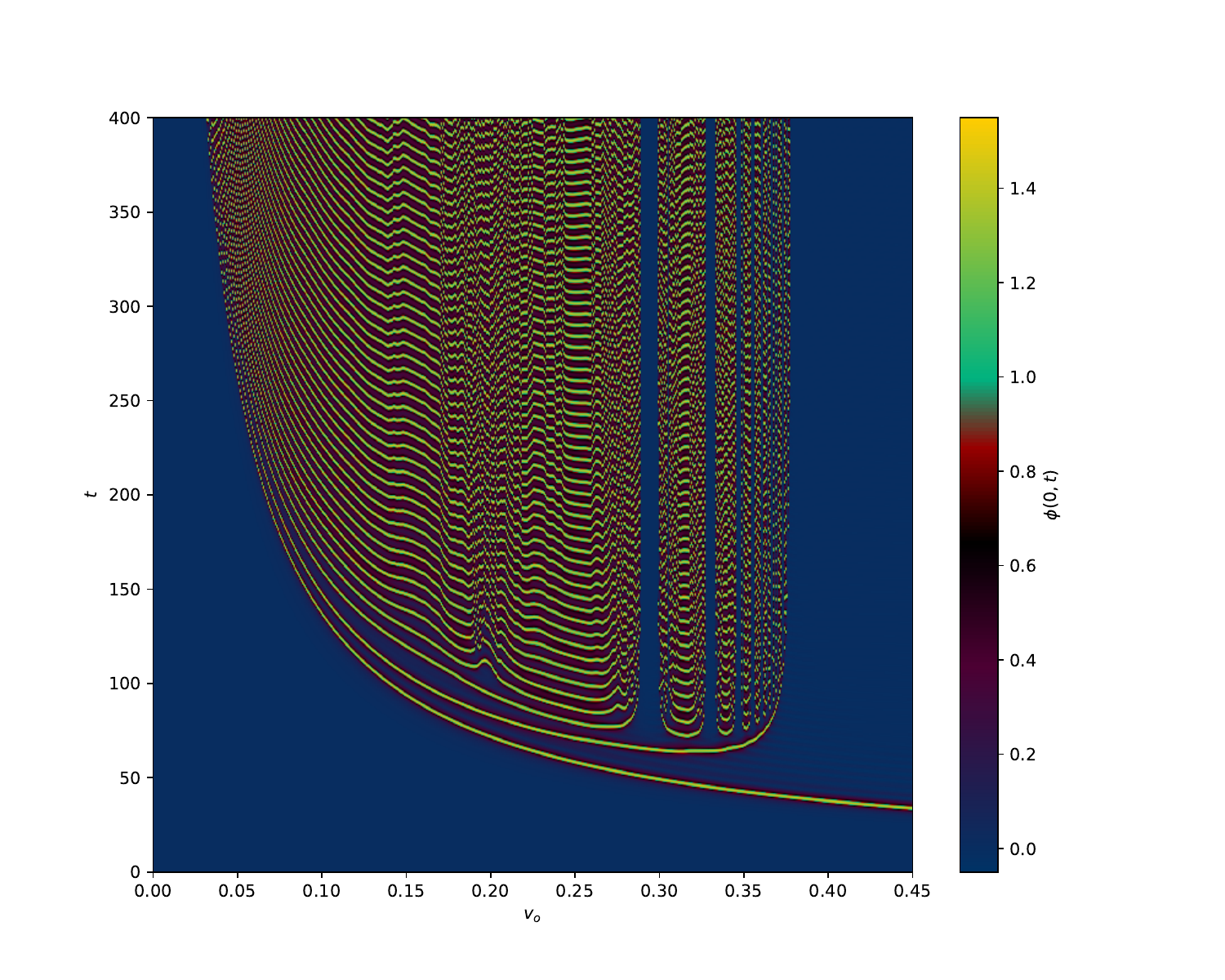}}
    \caption{Field at the collision center as a function of time and $v_0$. Kink-antikink collisions are depicted in (a) and antikink-kink collisions in (b). Parameters are $A^2=C^2=8/3$.}
 \label{fig:COM10}
\end{figure}

In Fig.~\ref{fig:COM10}, we depict the field at the collision center as a function of time and initial velocities $v_0$. The kink-antikink ($K\bar{K}$) scenario is on the left, and the antikink-kink on the right ($\bar{K}K$). The color map is interpreted as follows. Each bounce can be seen as an abrupt change of color in (a) and marked lines in (b). If separation occurs, the color reaches a fixed value when followed vertically. The reflection region is characterized by a large range of $v_0$ where the field acquires a dark blue color after a single bounce and may contain small two-bounce regions accumulating on its edge.

The antikink-kink scattering output corresponds precisely to the collisions obtained in Ref.~\cite{Campos2020} because the stability potential in the two models coincides, and also there is no sector switching. Kink-antikink collisions, on the other hand, exhibit sector switching, so they do not behave as the model designed in Ref.~\cite{Campos2020}, even for symmetric parameters. The simulations show that the resonance windows are narrower compared to the antikink-kink case, a pattern that will be observed in subsequent sections. For our toy model, kink-antikink and antikink-kink collisions behave identically at the linear level. They both have the same stability potential. In this sense, we have constructed a setup where the only difference between the two is whether sector switching is allowed. Therefore, the narrower windows for kink scattering with sector switching suggest that resonance windows may be less robust in this scenario.

According to theoretical calculations, the vibrational mode frequency is equal to $\omega_T = 1.533$. On the other hand, the numerically measured frequencies are $\omega_{\bar{K}K} = 1.500$ and $\omega_{K\bar{K}}= 1.499$. As expected, the theoretical and numerical frequencies agree closely, with only a two percent relative error. It is important to mention that small differences between measured and theoretical values are expected due to the complexity of the phenomenon we are modeling.

\subsection{$(A^2, C^2)=(1.5, 7)$ and $(A^2, C^2)=(1.1, 5.5)$}
	
Now, we can consider two cases, both without a localized mode but containing delocalized modes. They can be achieved by fixing $A^2 = 1.5$ and $C^2 = 7$ for the first scenario and $A^2 = 1.1$ and $C^2 = 5.5$ for the second one. Here, we construct resonance windows via the excitation of delocalized modes, first described in \cite{Dorey2011Kink} for the $\phi^6$ model.

\begin{figure}
    \centering
    \subfigure[]{\includegraphics[width=0.48\textwidth]{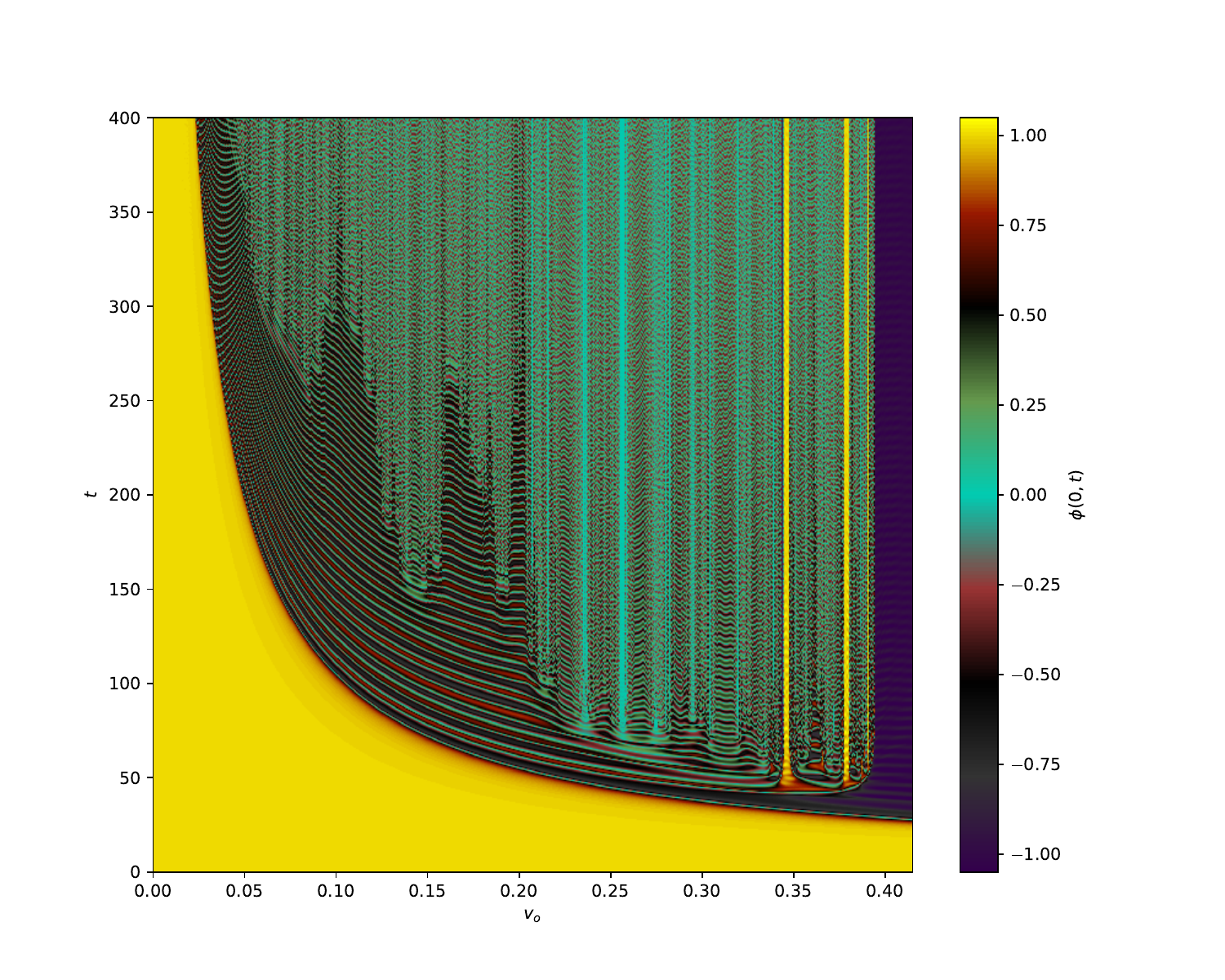}}
    \subfigure[]{\includegraphics[width=0.48\textwidth]{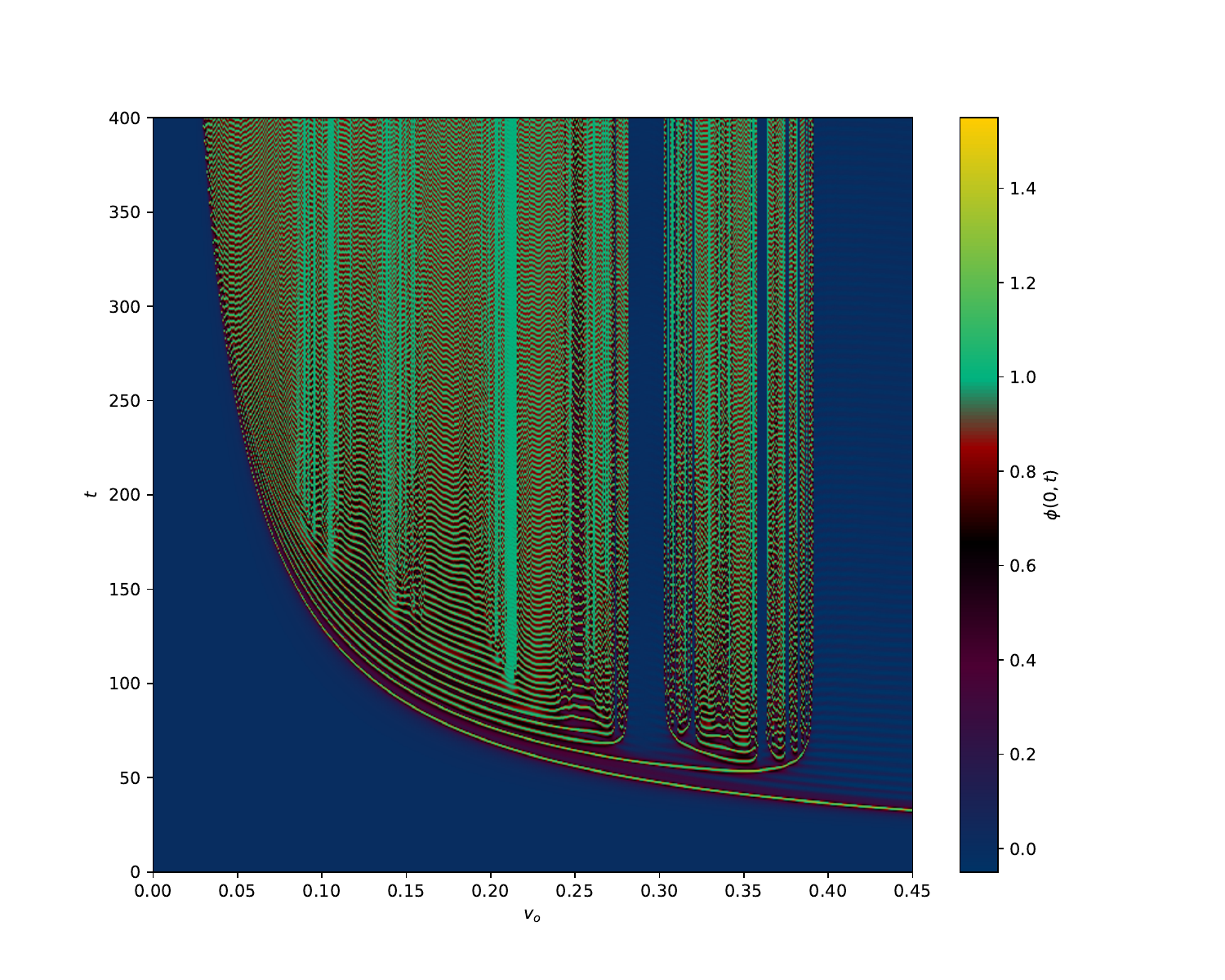}}  
    \caption{Field at the collision center as a function of time and $v_0$. Kink-antikink collisions are depicted in (a) and antikink-kink collisions in (b). Parameters are $A^2=1.5$ and $C^2=7$.}
    \label{fig:COL01}
\end{figure}

\begin{figure}
    \centering
    \subfigure[]{\includegraphics[width=0.48\textwidth]{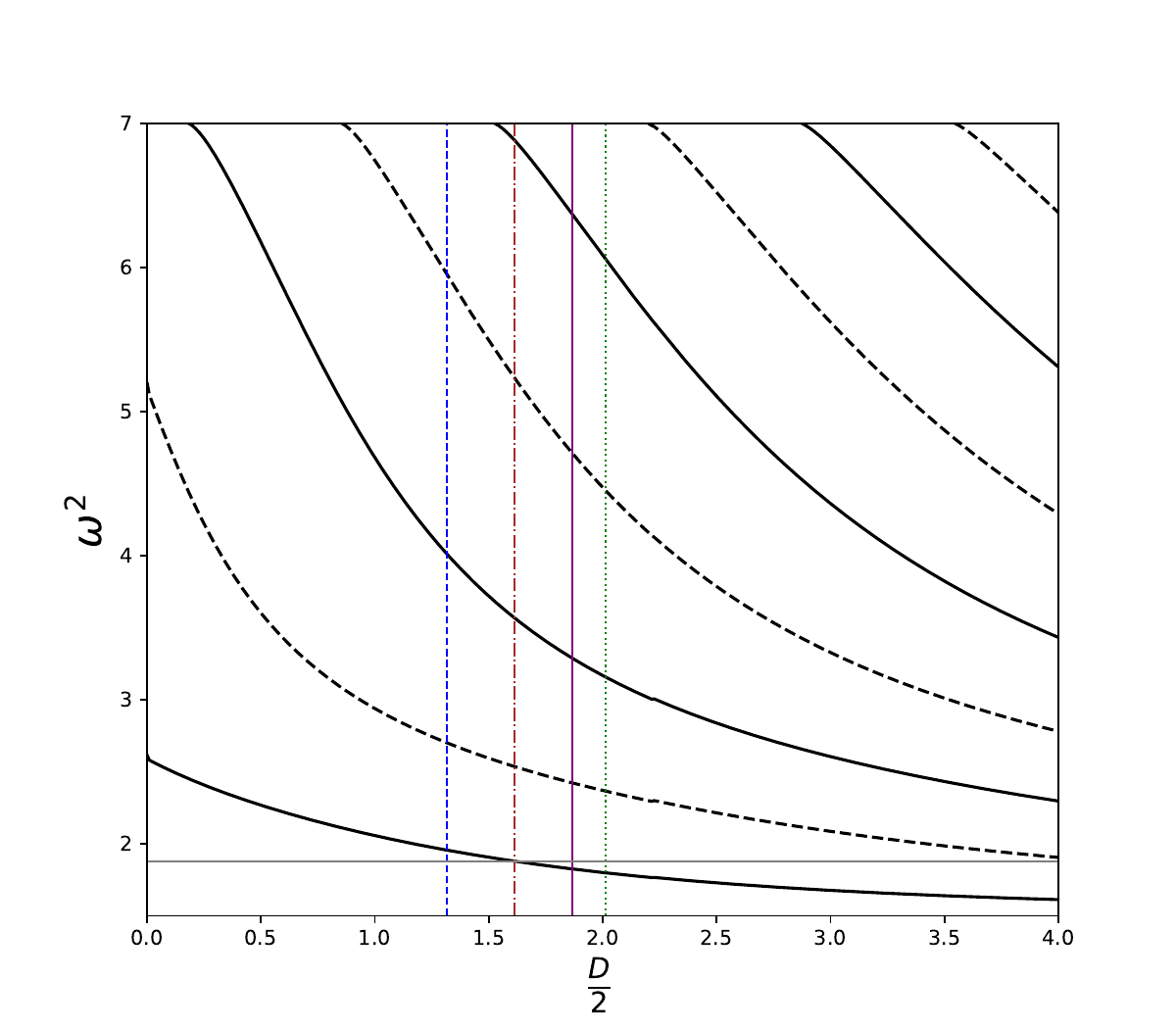}}
    \subfigure[]{\includegraphics[width=0.48\textwidth]{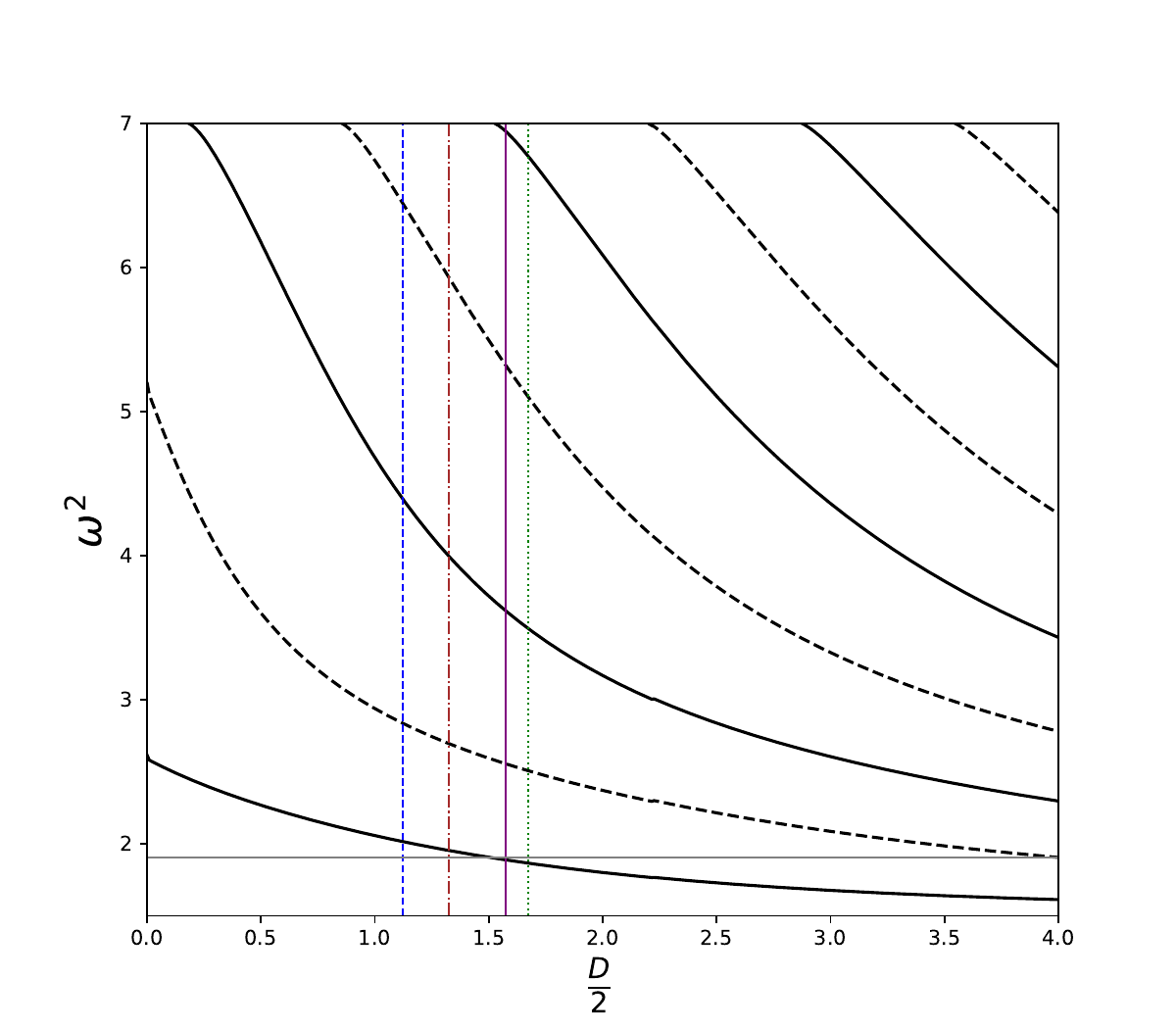}}  
    \caption{Spectrum of delocalized modes as a function of the interkink half-separation for (a) kink-antikink and (b) antikink-kink configurations. Parameters are $A^2=1.5$ and $C^2=7$. Delocalized modes exist only in the exhibited range, which is $A^2<\omega^2<C^2$.}
    \label{fig:spec01}
\end{figure}

In Fig.~\ref{fig:COL01}, we show the field at the collision center as a function of time and initial velocity for $A^2 = 1.5$ and $C^2 = 7$. In both cases, resonance windows are observed, although narrower in the kink-antikink scenario as before. Our toy model not only exhibits the resonant structure expected for antikink-kink collisions but also reveals a resonant structure in kink-antikink collisions, provided that the meson masses of the two vacua are swapped. The resonant frequency obtained from the numerical simulations are $\omega_{K\bar{K}} = 1.371$ and $\omega_{\bar{K}K} = 1.381$. To evaluate whether such values agree with the resonant energy exchange mechanism via the excitation of delocalized modes, we present the analytical delocalized modes frequencies as a function of the separation between the kinks in Fig.~\ref{fig:spec01}. As vertical lines, we plot the maximum distance between the kinks in the numerical simulations of the first four bounce windows. The horizontal lines mark the measured frequency. The lowest is the only delocalized mode compatible with the measured frequency in the range delimited by the vertical lines. Therefore, our results agree with the current understanding of the phenomenon.

\begin{figure}
    \centering
    \subfigure[]{\includegraphics[width=0.48\textwidth]{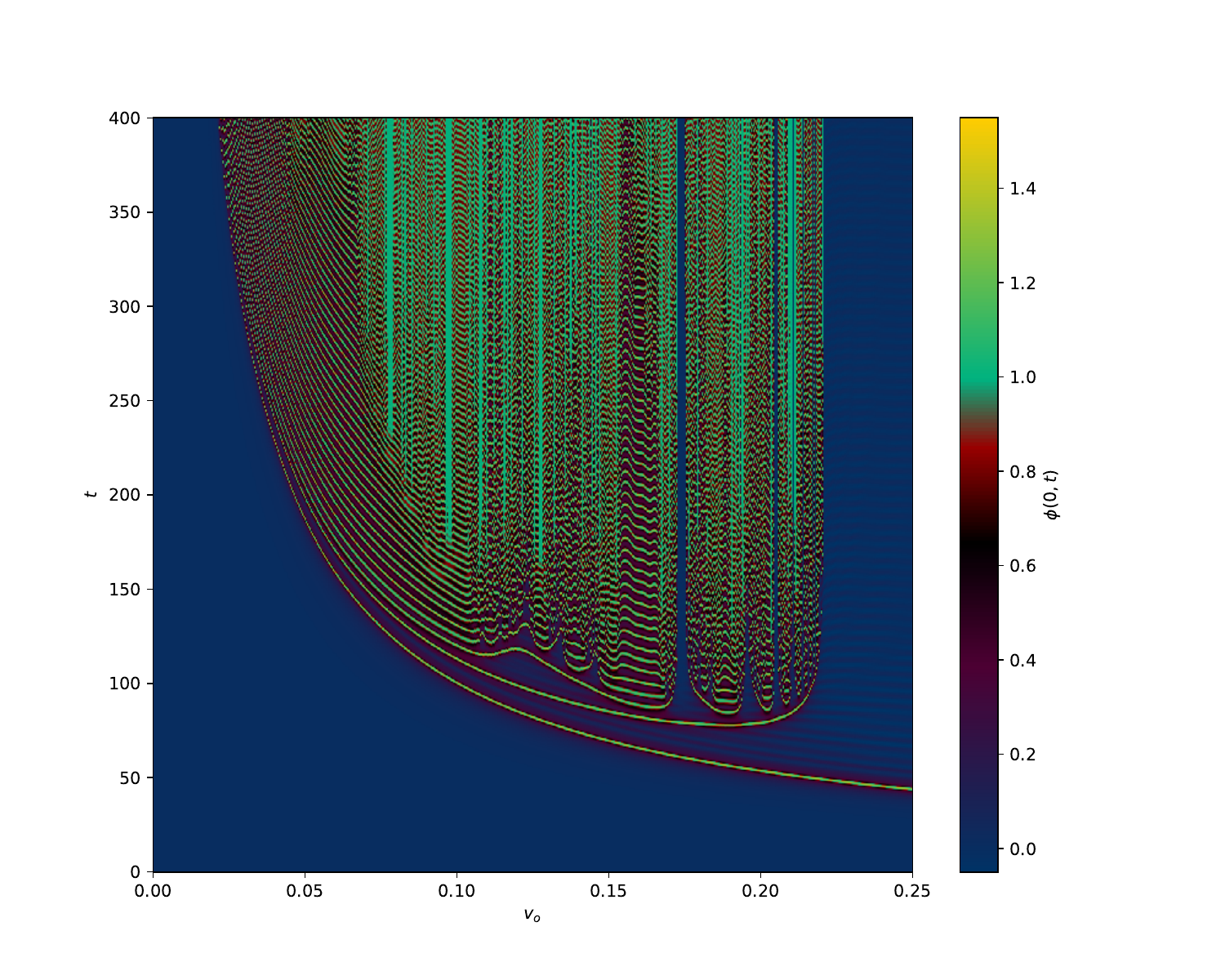}}
    \subfigure[]{\includegraphics[width=0.48\textwidth]{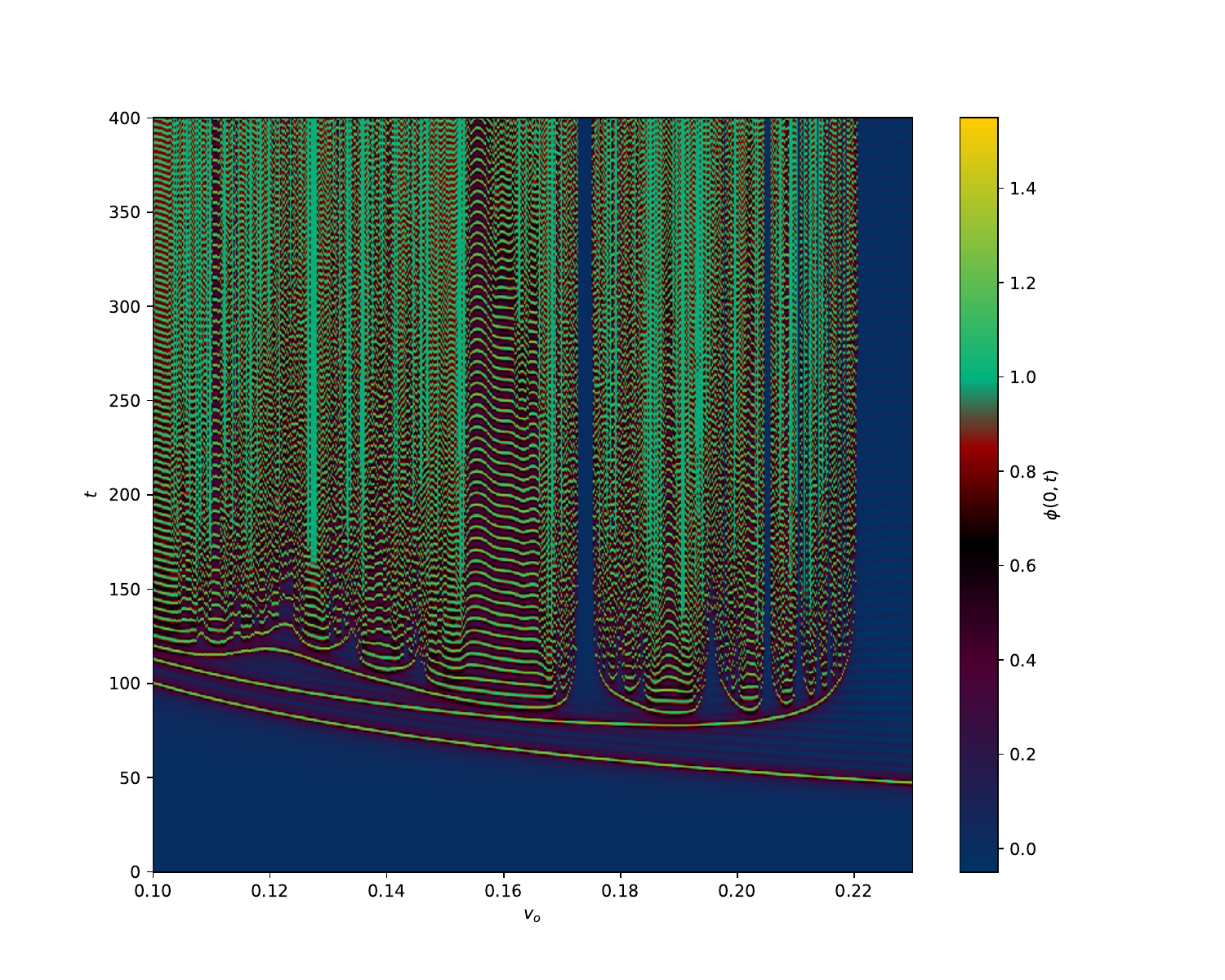}}  
    \caption{Field at the collision center as a function of time and $v_0$. Antikink-kink collisions are depicted in (a) and the zoom in (b). Parameters are $A^2=1.1$ and $C^2=5.5$.}
    \label{fig:COL0n}
\end{figure}

A notable feature of the $\phi^6$ model is the appearance of not only resonance windows, but also false ones. By adjusting the parameters $A$ and $C$, we can investigate if our toy model can reveal the same feature. In Fig.~\ref{fig:COL0n}, considering $A^2 = 1.1$ and $C^2 = 5.5$, a mixed structure of false and true windows appears. In \cite{Dorey2011Kink}, the authors argued that false resonance windows are the result of competition between the lowest delocalized mode and higher frequency ones. Here, we can also associate the appearance of false windows with the change in the spectrum from the first scenario, $A^2 = 1.5$ and $C^2 = 7$.

\subsection{$(A^2, C^2)=(2.5, 6.125)$}
	
\begin{figure}
    \centering
    \subfigure[]{\includegraphics[width=0.48\textwidth]{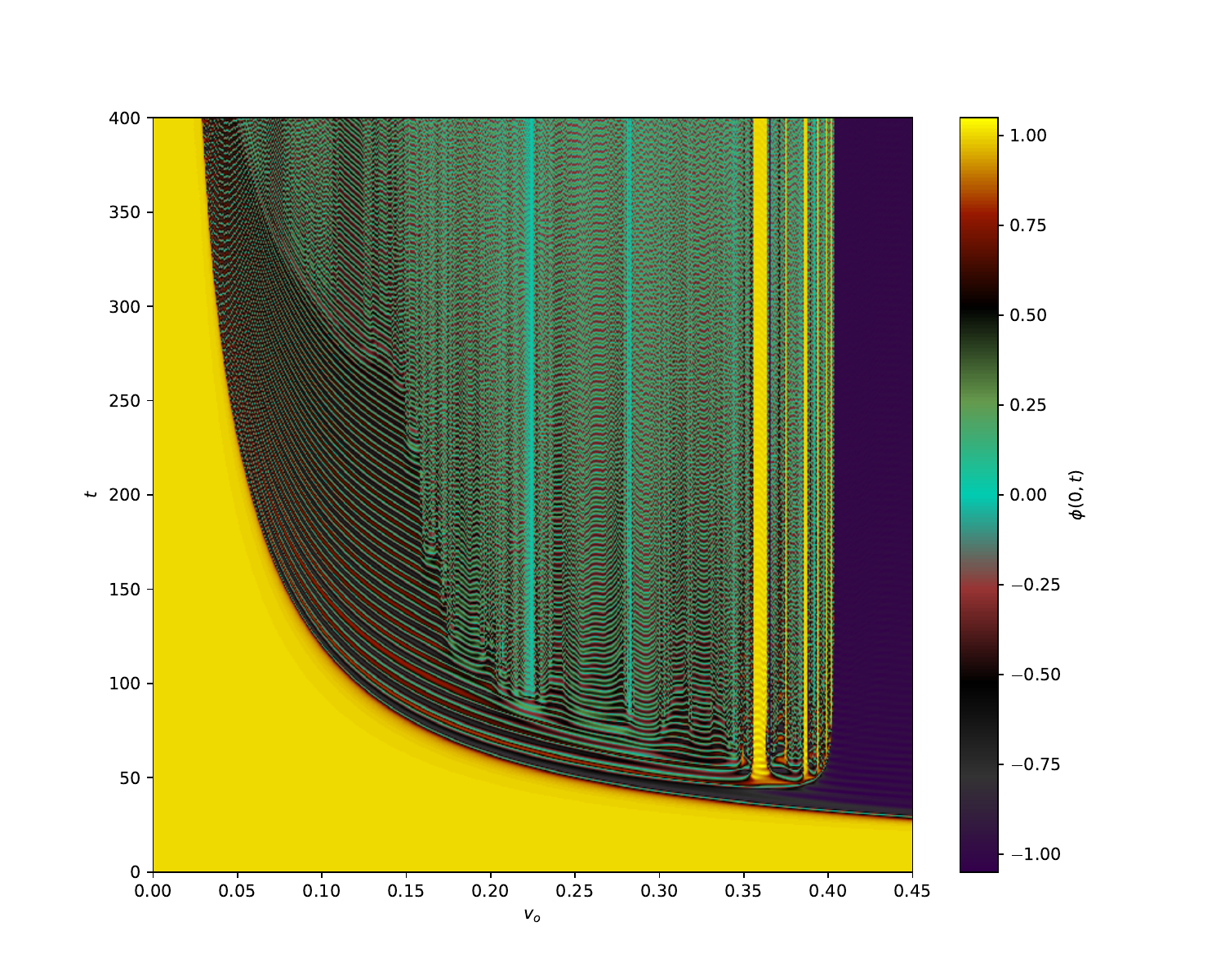}}
    \subfigure[]{\includegraphics[width=0.48\textwidth]{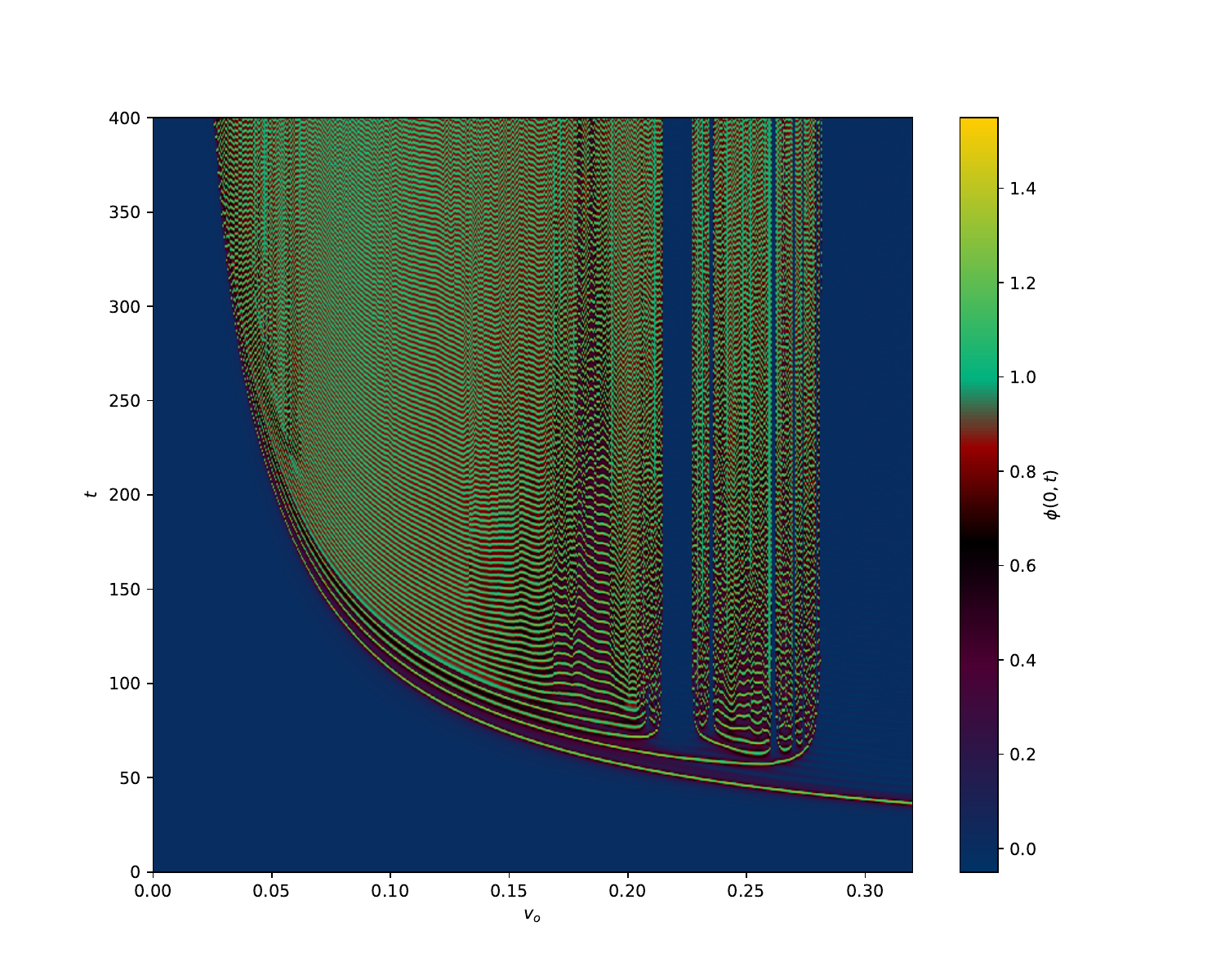}}  
    \caption{Field at the collision center as a function of time and $v_0$. Kink-antikink collisions are depicted in (a) and antikink-kink collisions in (b). Parameters are $A^2=2.5$ and $C^2=6.125$.}
 \label{fig:COL11}
\end{figure}

Let us also investigate the scenario containing one localized as well as some delocalized modes by fixing the parameters $A^2 = 2.5$ and $C^2 = 6.125$. The scattering output is summarized in Fig.~\ref{fig:COL11}. As before, resonance windows are observed, with the kink-antikink case exhibiting narrower windows.

\begin{figure}
    \centering
    \subfigure[]{\includegraphics[width=0.48\textwidth]{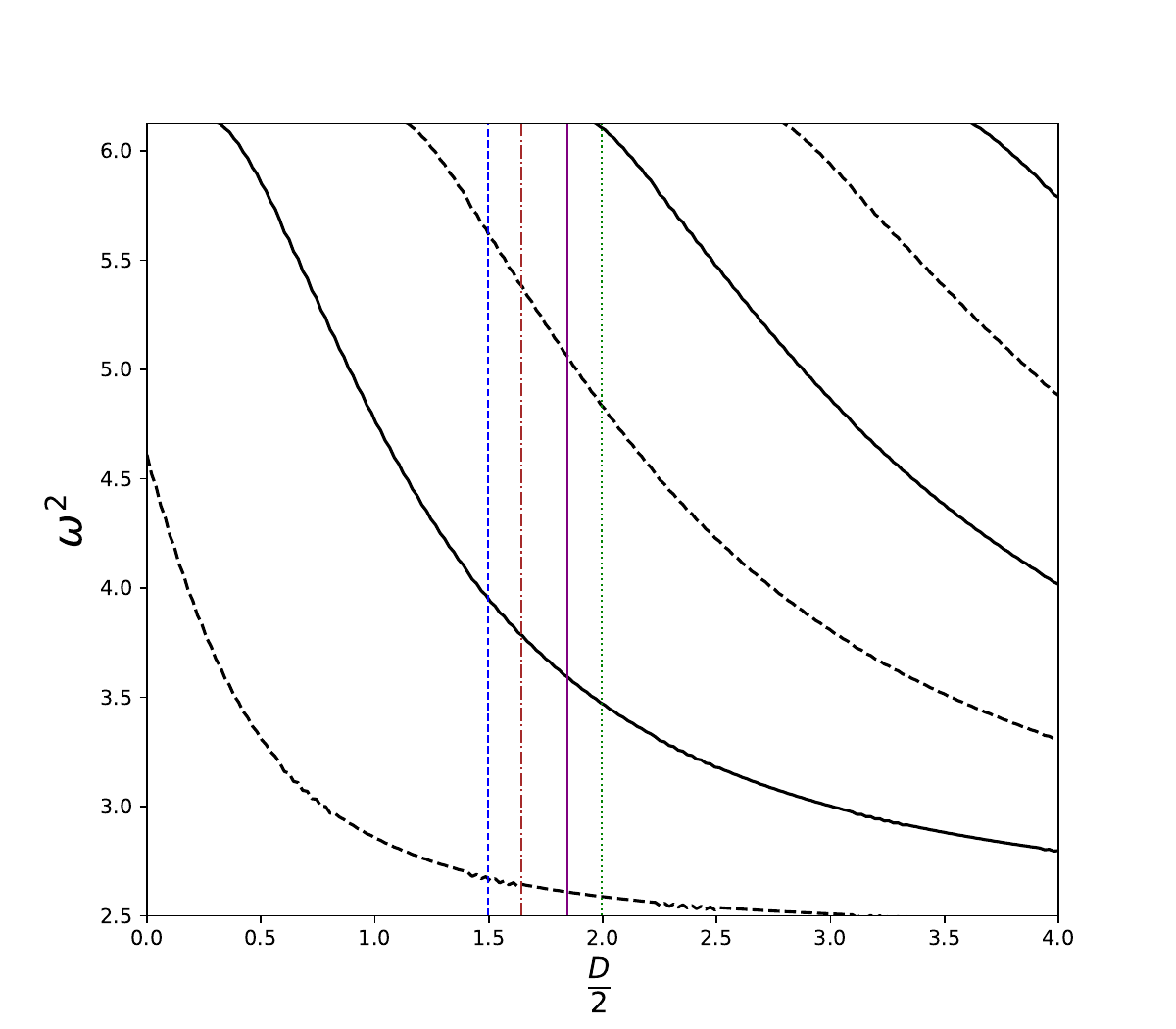}}
    \subfigure[]{\includegraphics[width=0.48\textwidth]{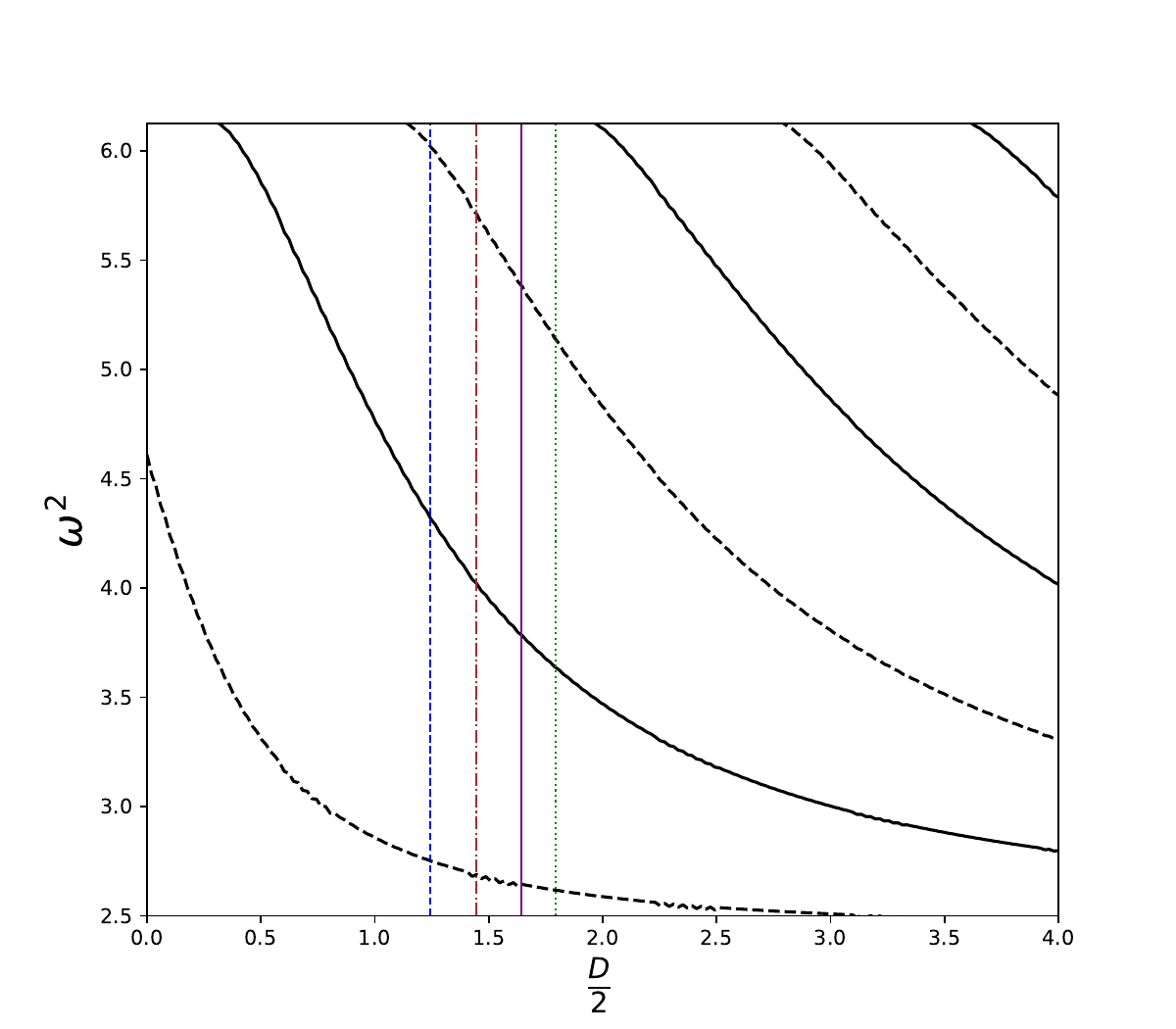}}  
    \caption{Spectrum of delocalized modes as a function of the interkink half-separation for (a) kink-antikink and (b) antikink-kink configurations. Parameters are $A^2=2.5$ and $C^2=6.125$. Delocalized modes exist only in the exhibited range, which is $A^2<\omega^2<C^2$.}
 \label{fig:spec11}
\end{figure}

The theoretical frequency of the localized mode is $\omega_{T} = 1.563$. Furthermore, the frequency of delocalized mode as a function of $D/2$ is shown in Fig.~\ref{fig:spec11}. Recall that they obey $A^2 < \omega^2 < C^2$. The interkink distance for the first four bounce windows is also marked as vertical lines. The frequencies should be compared with the associated measured frequencies $\omega_{\bar{K}K} = 1.528$ and $\omega_{K\bar{K}} = 1.501$. For the localized frequency, the associated relative errors are $\delta \omega_{\bar{K}K} = 2.1\%$ and $\delta \omega_{K\bar{K}} = 4.0\%$. Although the delocalized modes begin at $A\simeq 1.581$, the odd delocalized modes cannot be excited due to the even symmetry of the initial conditions. Thus, the measured frequency should be compared to the frequency of the lowest even delocalized mode. These frequencies, however, are significantly different from the measured values. Therefore, the localized mode is responsible for the resonant structure, and the presence of a delocalized mode may be insufficient to disrupt the resonant behavior in this scenario.

\subsection{$(A^2, C^2)=(6.05, 6.05)$}
	
\label{sec:D}	
	
\begin{figure}
    \centering
    \subfigure[]{\includegraphics[width=0.48\textwidth]{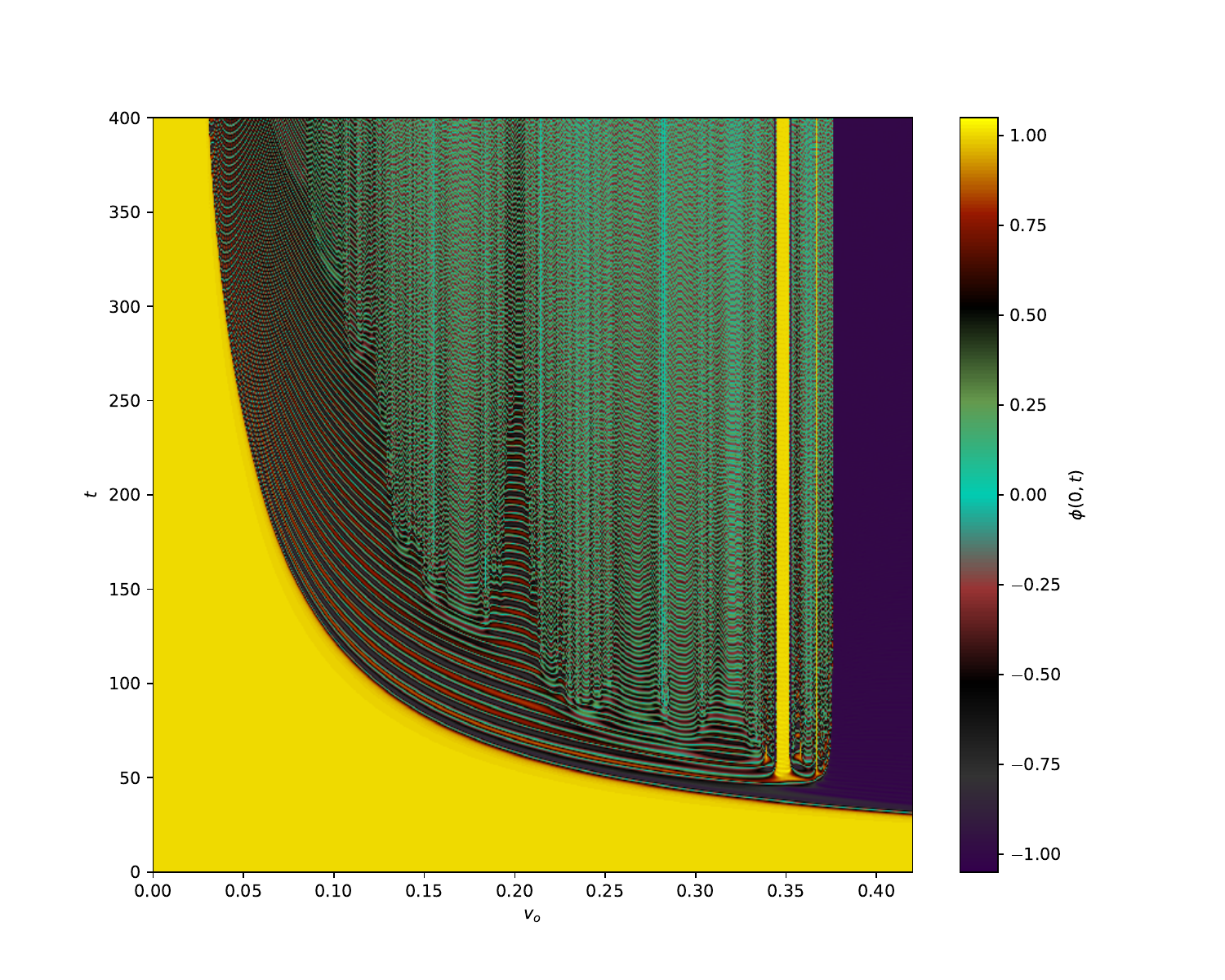}}
    \subfigure[]{\includegraphics[width=0.48\textwidth]{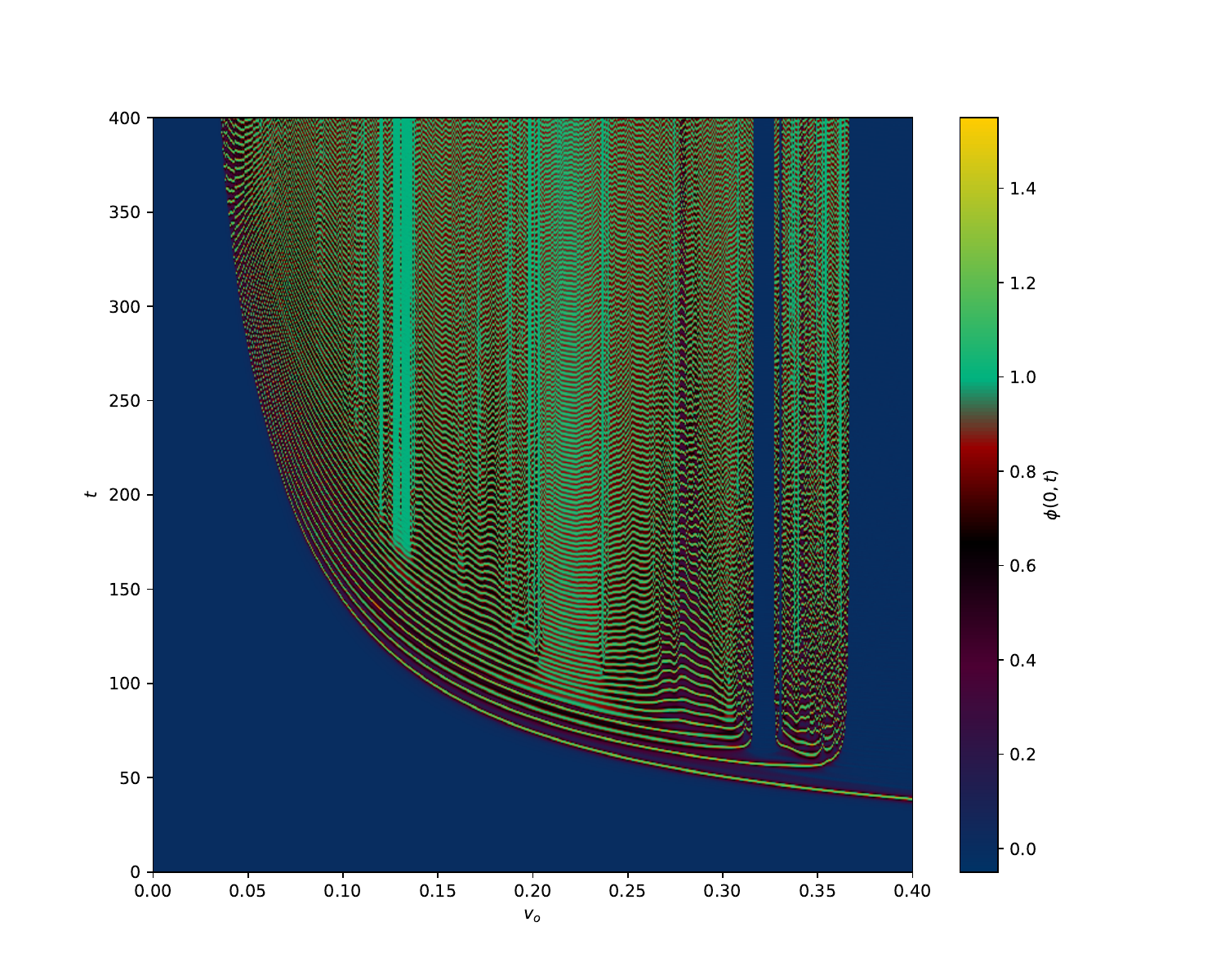}}  
    \caption{Field at the collision center as a function of time and $v_0$. Kink-antikink collisions are depicted in (a) and antikink-kink collisions in (b). Parameters are $A^2 = C^2 = 6.05$.}
 \label{fig:COM20}
\end{figure}

The present subsection explores the scenario with two localized and no delocalized modes. The parameters are symmetrically set to $A^2 = C^2 = 6.05$, leading to the following theoretical frequencies $\omega_{T_1} = 1.678$ and $\omega_{T_2} = 2.457$. The scattering output is summarized in Fig.~\ref{fig:COM20}, following the same format as previous subsections. In both scenarios, kink-antikink and antikink-kink collisions, the resonance windows are mostly suppressed. This result is consistent with the literature, which indicates that resonance windows are suppressed when more than one localized vibrational mode is present \cite{Simas2016Sup}. 
However, there are a few exceptions to this picture. For instance, resonance windows were identified in Ref.~\cite{Dorey2021Res} when two vibrational modes were present. The authors developed a model with a free parameter that increases the number of vibrational modes when adjusted. In such a model, the resonance windows exist when a second vibrational mode appears, but only within a narrow range of parameters near that point. Moreover, the authors in Ref.~\cite{Marjaneh2022Collisions} found resonance windows with more than a single localized mode.

Although we found very few resonance windows, we were able to locate many false resonance windows in the antikink-kink scenario. False resonance windows are intervals of initial velocities around local maxima in the time between bounces, where the kinks acquire a significant amount of energy from the vibrational mode at the second bounce but not enough to completely separate. We measured the corresponding resonant frequency as $\omega_{\bar{K}K} = 1.588$, with a relative error of $\delta \omega_{\bar{K}K} = 5.2 \%$ compared to the lowest vibrational mode. In the kink-antikink scenario, even false resonance windows were absent, consistent with the windows being narrower whenever present.

Therefore, our simulations confirm that resonance windows tend to be suppressed when two localized vibrational modes are present. Consistently with Refs.~\cite{Dorey2021Res, Marjaneh2022Collisions}, our results also suggest that this suppression occurs gradually rather than abruptly with the appearance of a second mode.

\subsection{$(A^2, C^2)=(1.5, 9.0)$}

\begin{figure}
    \centering
    \subfigure[]{\includegraphics[width=0.48\textwidth]{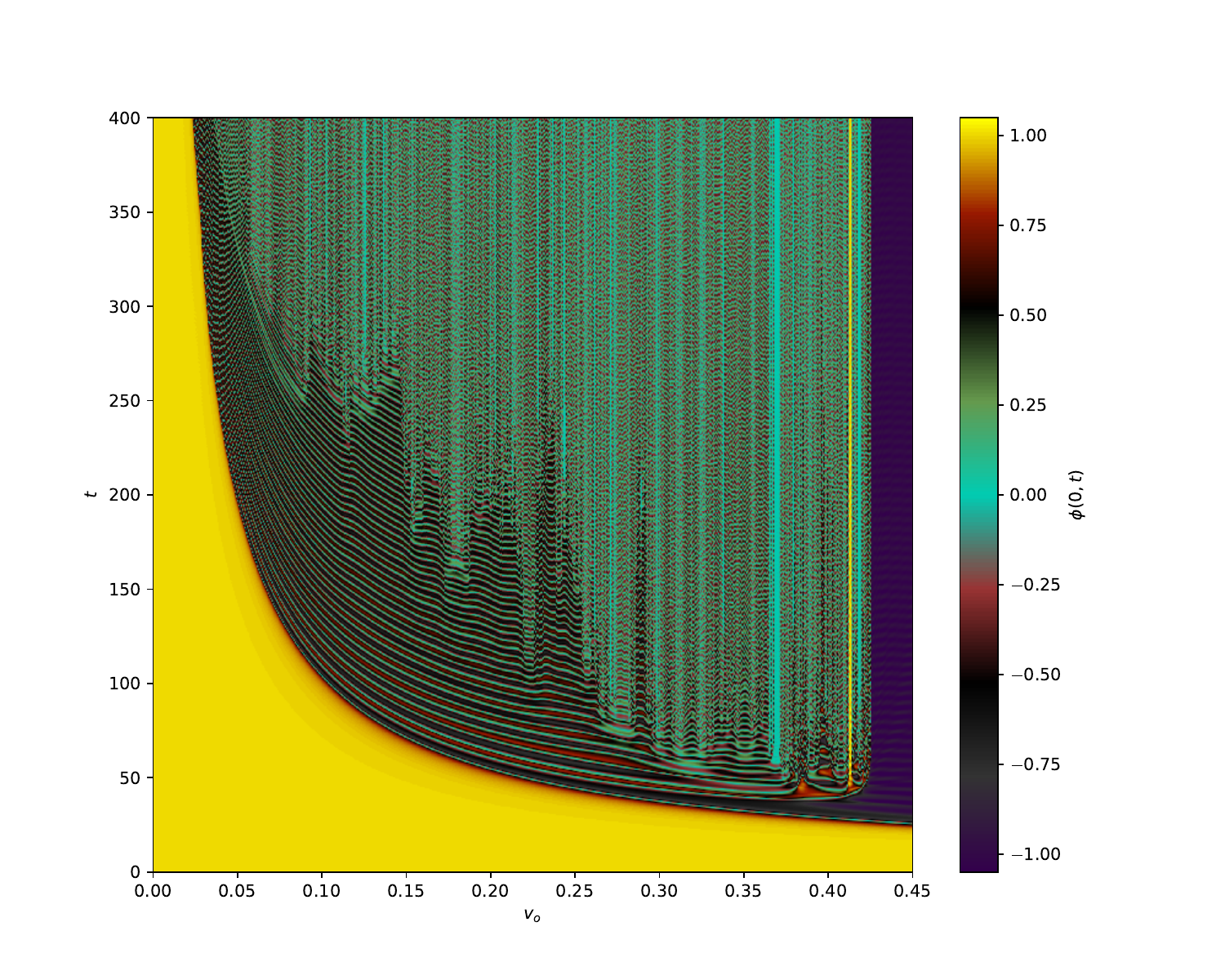}}
    \subfigure[]{\includegraphics[width=0.48\textwidth]{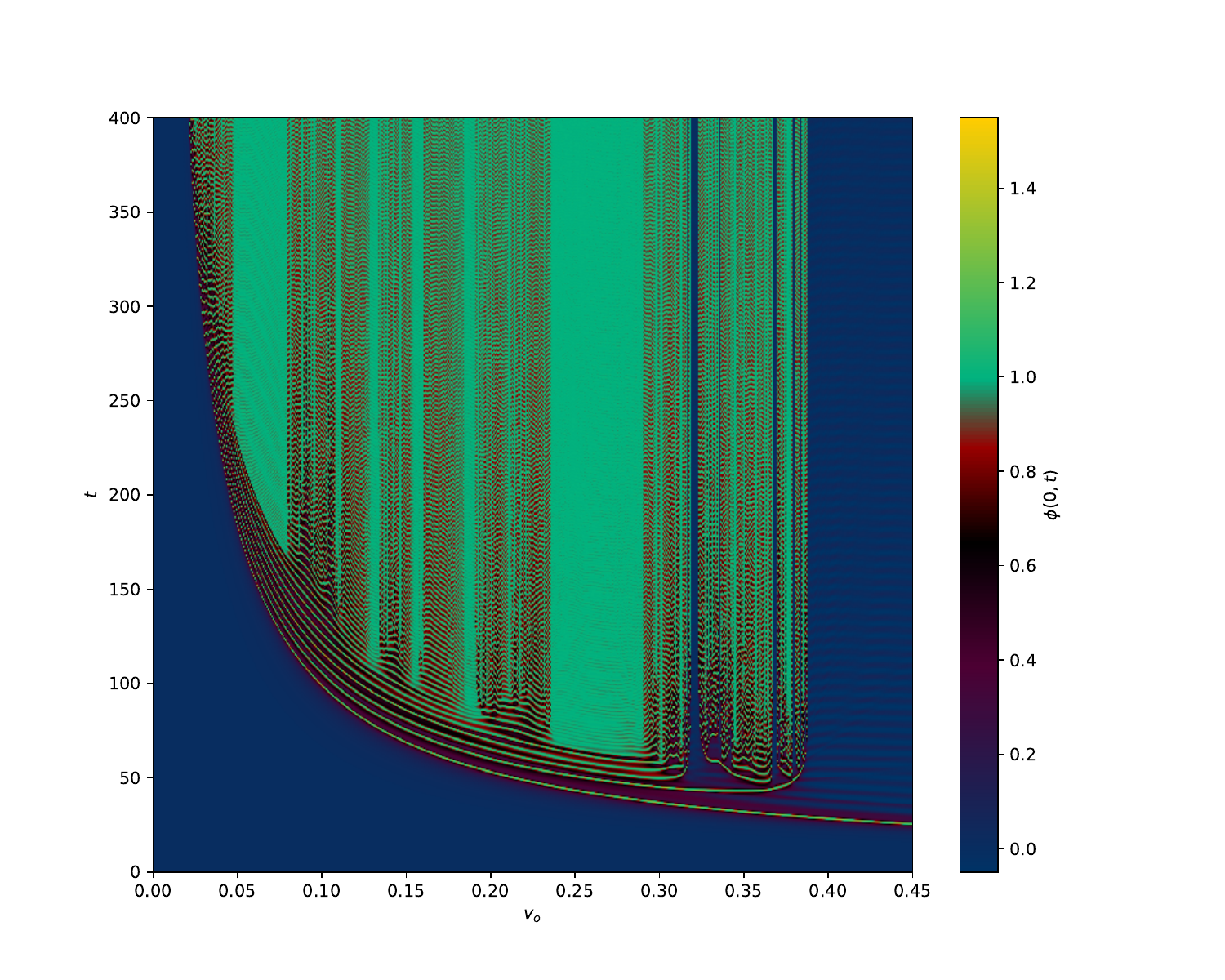}}  
    \caption{Field at the collision center as a function of time and $v_0$. Kink-antikink collisions are depicted in (a) and antikink-kink collisions in (b). Parameters are $A^2 = 1.5$ and $C^2 = 9.0$.}
 \label{fig:COM02}
\end{figure}

\begin{figure}
    \centering
    \subfigure[]{\includegraphics[width=0.48\textwidth]{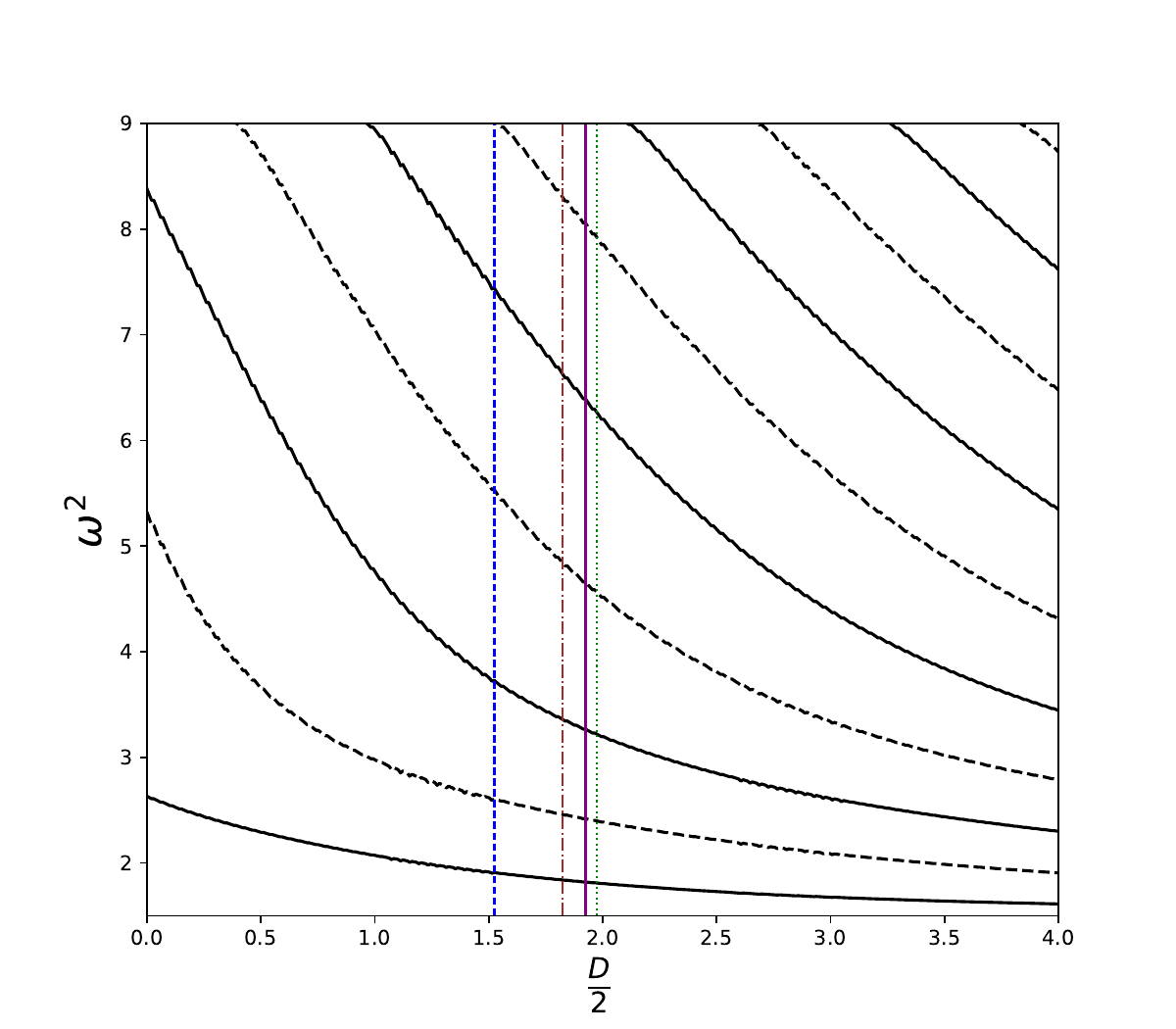}}
    \subfigure[]{\includegraphics[width=0.48\textwidth]{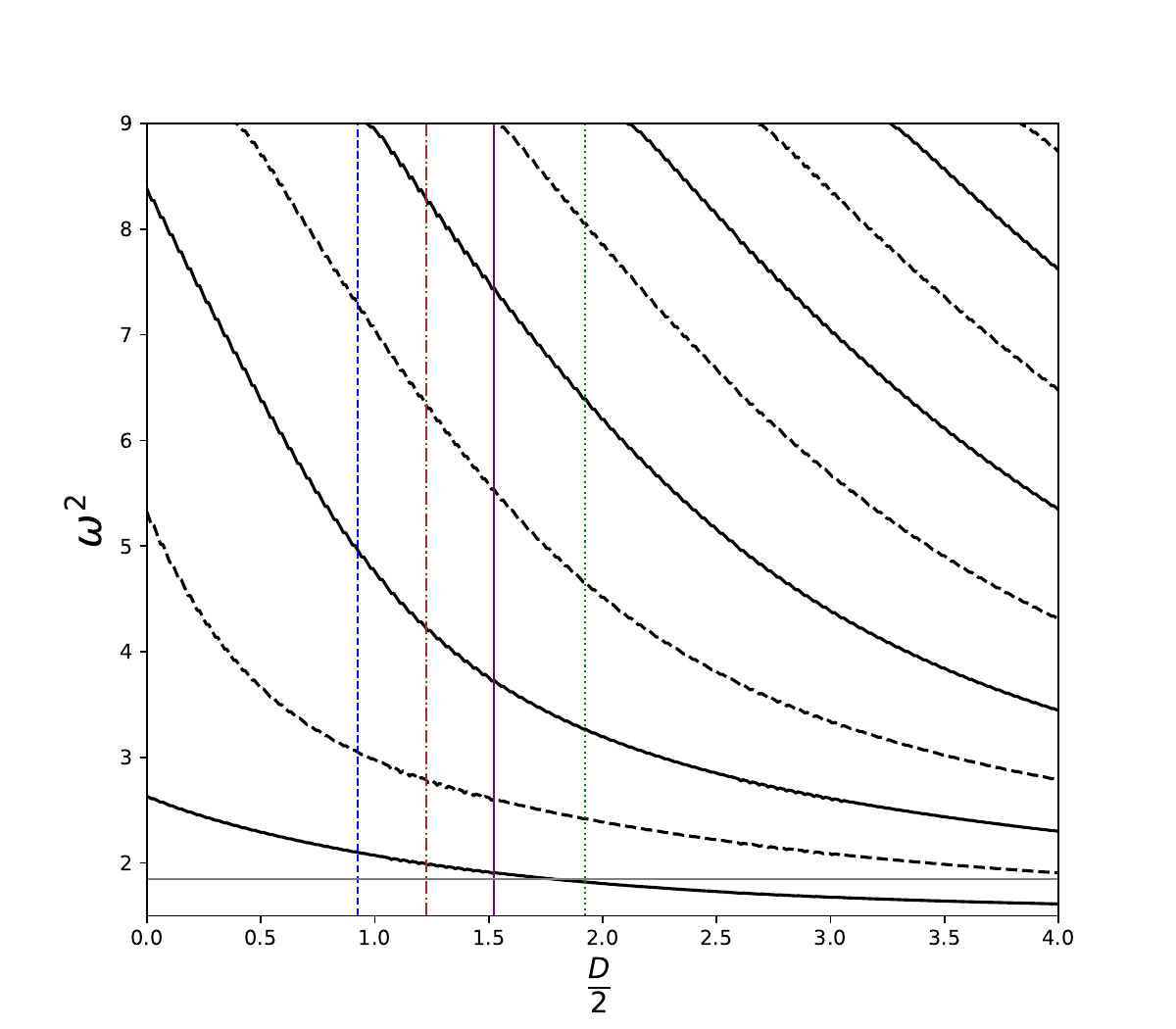}}  
    \caption{Spectrum of delocalized modes as a function of the interkink half-separation for (a) kink-antikink and (b) antikink-kink configurations. Parameters are $A^2=1.5$ and $C^2=9.0$. Delocalized modes exist only in the exhibited range, which is $A^2<\omega^2<C^2$.}
 \label{fig:spec02}
\end{figure}

The next scenario contains parameters $A^2 = 1.5$ and $C^2 = 9.0$ with no localized mode. It leads to a larger set of delocalized modes for typical kink-antikink (and antikink-kink) separations between bounces. The scattering output is depicted in Fig.~\ref{fig:COM02} and the delocalized mode frequencies as a function of $D/2$ is shown in Fig.~\ref{fig:spec02}. Both are in the same format as before.

Somewhat surprisingly, the windows are mostly suppressed in the kink-antikink case. As the windows are more fragile in this scenario, they were suppressed by the presence of extra delocalized modes. Therefore, we conclude that windows suppression may also occur via the presence of delocalized modes. For antikink-kink collisions, we identify resonance windows, as is generally expected. The measured frequency is $\omega_{\bar{K}K} = 1.358$. Once more, we measure the largest interkink distance at the first four resonance windows. In that range of interkink distances, the numerical frequency is indeed compatible with the lowest delocalized mode.

\subsection{$(A^2, C^2)=(7.0,9.5)$ and $(A^2, C^2)=(7.0,12.0)$}

\begin{figure}
    \centering
    \subfigure[]{\includegraphics[width=0.48\textwidth]{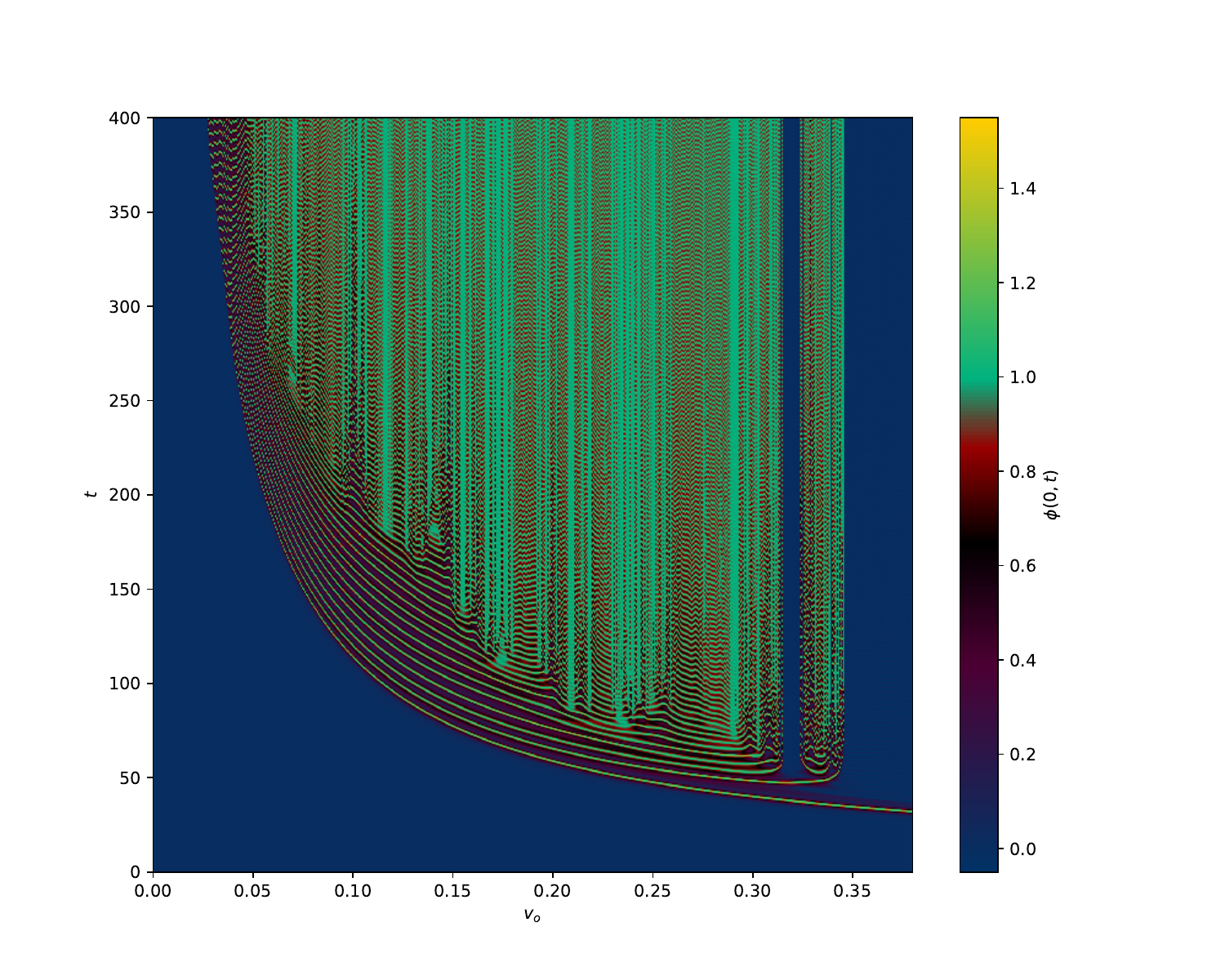}}  
    \subfigure[]{\includegraphics[width=0.48\textwidth]{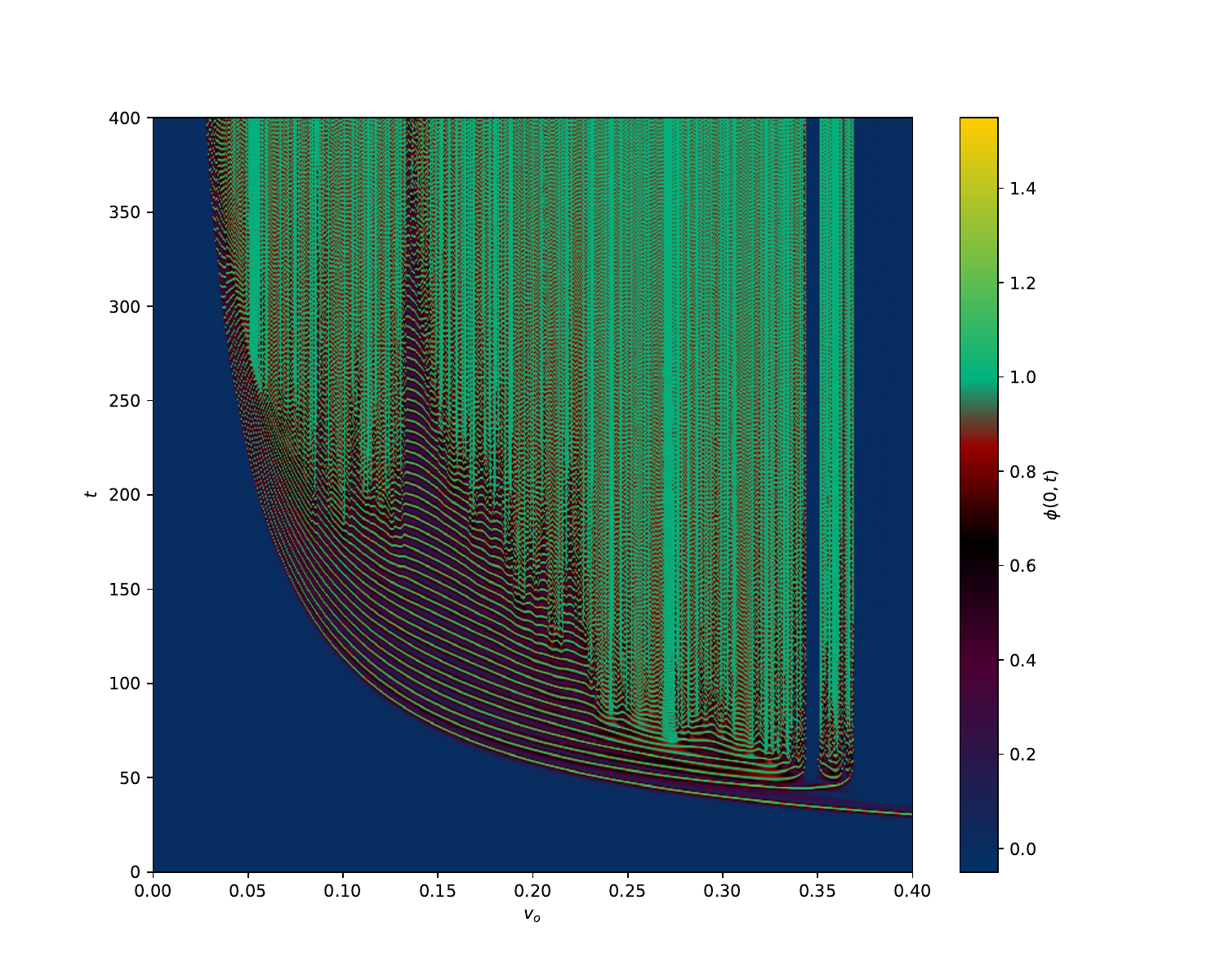}}
    \caption{Field at the collision center as a function of time and $v_0$. We consider antikink-kink collisions with parameters (a) $(A^2,C^2)=(7.0,9.5)$ and (b) $(A^2,C^2) = (7.0,12.0)$.}
 \label{fig:COM2122}
\end{figure}

In this section, we consider the cases with parameters $(A^2,C^2)=(7.0,9.5)$ and $(A^2,C^2) = (7.0,12.0)$. Both contain a set of two localized modes as well as some delocalized modes, with the second case containing a larger set. For kink-antikink collisions, resonance windows are mostly suppressed. Such a result is expected, given that they were also absent for the case with two localized modes shown in sec.~\ref{sec:D}. The remaining scattering output, corresponding to antikink-kink collisions, is shown in Fig.~\ref{fig:COM2122}. 

For $(A^2, C^2)=(7.0,9.5)$, the theoretical localized frequencies are $\omega_{T_1} = 1.697$, $\omega_{T_2} = 2.624$ and the delocalized modes start to appear above $\sqrt{7}$, i.e., approximately 2.646. We were able to find a few resonance windows and many false ones. The numerically measured frequency is $\omega_{\bar{K}K} = 1.563$. Compared with the lowest localized mode, the relative error equals $\delta \omega_{\bar{K}K} = 7\%$. At this point, it is important to emphasize a general feature of kink-antikink (and antikink-kink) collisions; the measured frequency corresponds to the lowest vibrational mode available whenever resonance windows are present, whether true or false.

For $(A^2, C^2)=(7.0,12.0)$, just a few resonance windows are present. Moreover, we were not able to locate false resonance windows, indicating that the resonant structure is mostly suppressed due to the energy exchange with the extra modes.

\subsection{$(A^2, C^2)=(2.7,9.5)$}

\begin{figure}
    \centering
    \subfigure[]{\includegraphics[width=0.48\textwidth]{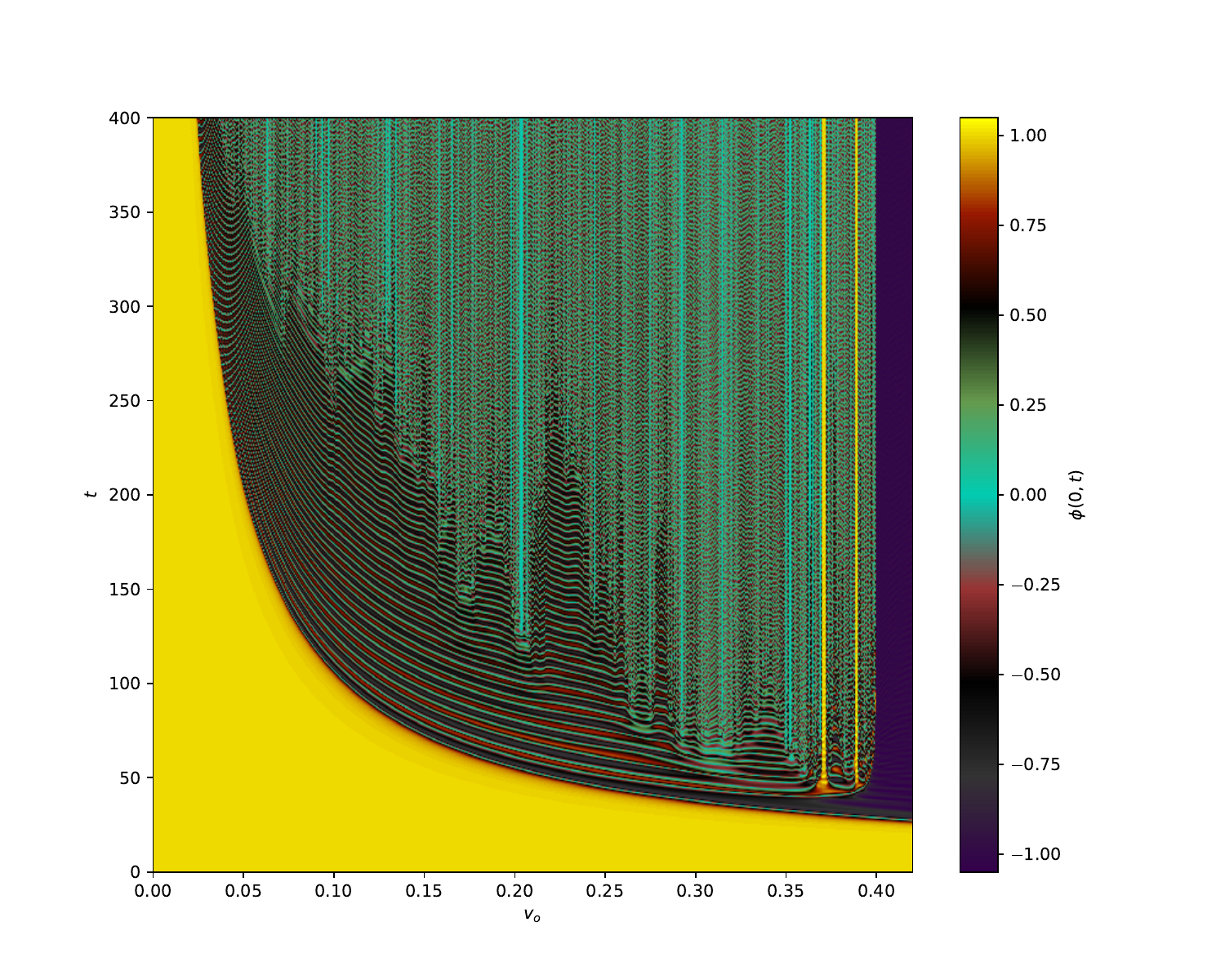}}
    \subfigure[]{\includegraphics[width=0.48\textwidth]{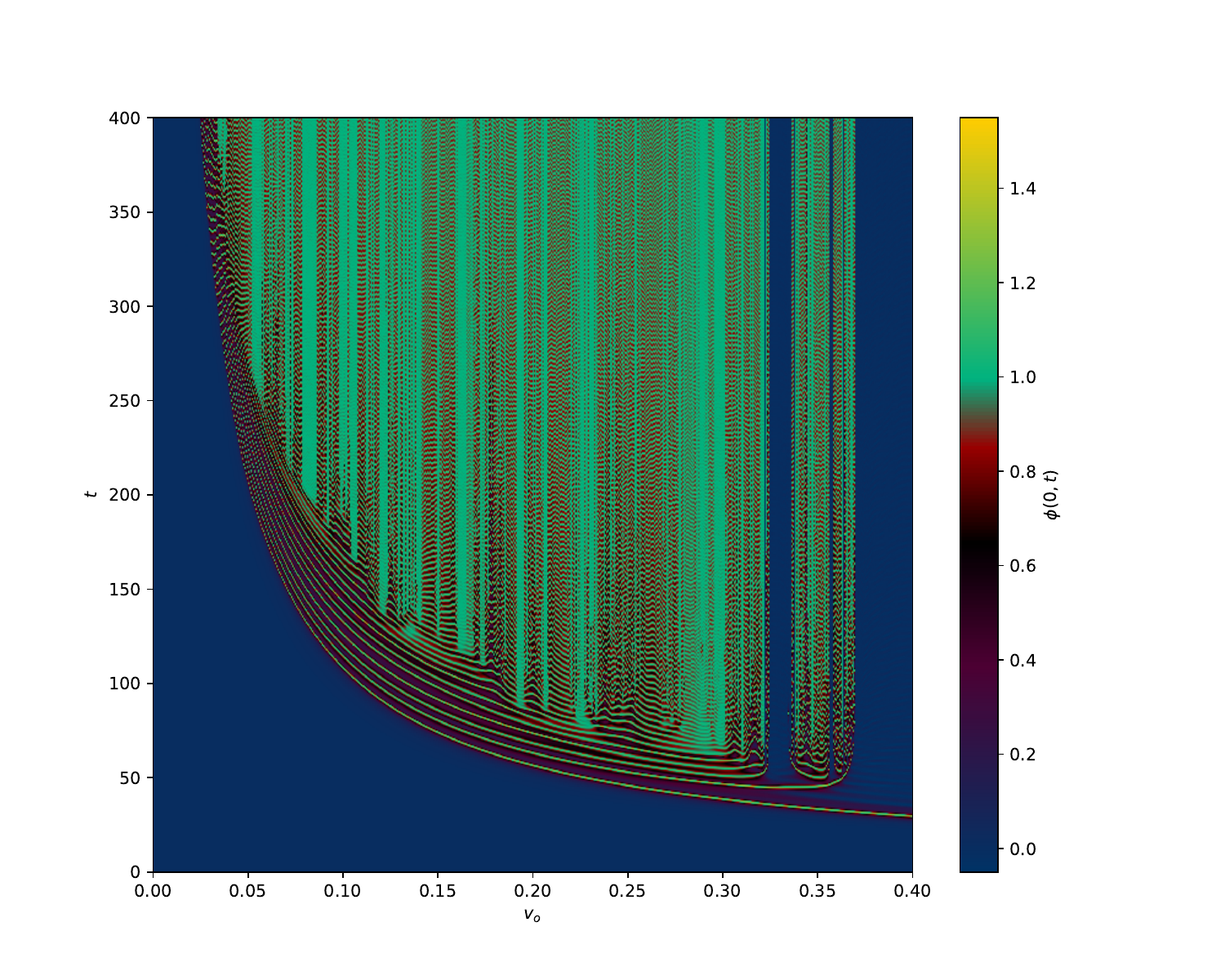}}  
    \caption{Field at the collision center as a function of time and $v_0$. Kink-antikink collisions are depicted in (a) and antikink-kink collisions in (b). Parameters are $A^2 = 2.7$ and $C^2 = 9.5$.}
 \label{fig:COM12}
\end{figure}

Finally, we fix $A^2 = 2.7$ and $C^2 = 9.5$. It corresponds to another scenario where the single kink contains a localized mode, and kink pairs contain some delocalized modes. The usual scattering output is shown in Fig.~\ref{fig:COM12}. In the kink-antikink scenario, resonance windows are absent. On the one hand, such a result is expected because the resonance windows are more fragile than the ones in the antikink-kink case. On the other hand, we observe that delocalized modes can surprisingly suppress resonance windows generated by localized modes. 

For antikink-kink collisions, the shape mode frequency is $\omega_{T} = 1.596$, while the delocalized modes start to appear above $\sqrt{2.7}$, i.e., approximately 1.643. The resonant structure in these collisions persists and is characterized by a frequency of $\omega_{\bar{K}K} = 1.548$. The error relative to the lowest vibrational mode is $\delta \omega_{\bar{K}K} = 3\%$. On the other hand, the lowest even delocalized mode frequency is quite far from the numerically measured frequency. Therefore, our results reinforce the notion that the lowest vibrational mode primarily governs the appearance of resonance windows.

\section{Conclusion}

In this study, we designed a piecewise $\phi^2$ toy model with a triple-well potential. Such construction leads to an asymmetric square-well potential for linearized perturbations around the kink solutions. Consequently, we could manipulate the number of localized and delocalized modes in kink collisions by adjusting the model's free parameters, $A$ and $C$. Furthermore, the frequencies were analytically determined using transcendental equations.

We explored many spectral scenarios for kink-antikink and antikink-kink collisions, with and without sector changing, respectively. Some of them were previously unexplored in the literature. Resonant behavior was observed in certain cases, while it was mostly suppressed in others. Across all scenarios studied, sector-changing collisions exhibited narrower resonance windows. Notably, the resonant frequency consistently aligned with the lowest available vibrational mode. 

For delocalized resonant behavior, the numerically measured delocalized frequency consistently matched with the lowest delocalized mode in the range of measured interkink distance. As the value of the largest interkink distance varied significantly between different bounce windows, it reinforces the necessity of constructing a collective coordinate model with dynamical delocalized modes rather than frozen ones to achieve a more realistic picture of resonant behavior from delocalized modes.
	
Collisions involving a single vibrational mode, whether localized or delocalized, exhibited a rich structure of resonance windows. Adding extra modes, even delocalized ones, can partially or entirely suppress this resonant structure. Specifically, resonance windows are narrower and more easily suppressed in kink-antikink collisions. In contrast, the antikink-kink scenario is more robust and remains unaffected by additional delocalized modes in certain cases. Other combinations are possible, as varying the parameters $A$ and $C$ allows for a wide range of scenarios. 

An interesting continuation of the present work is to analyze differences in radiation production between kink-antikink collisions with and without sector switching. Our toy model offers a perfect setup to analyze this issue, given that we can choose between allowing sector switching or not while keeping all linear features of the model fixed. To the best of our knowledge, it is the only model in the literature exhibiting such a feature. This could be achieved by measuring the amount of radiation that reaches the boundary at infinity after the kinks interact.

According to the standard resonance energy exchange mechanism, resonance windows appear because translational energy is converted into vibrational energy and vice-versa whenever the kinks reemerge after multiple bounces. In this work, we drew a more general picture, namely that energy is redistributed among all possible modes at each bounce. The frequent occurrence of resonance windows suggests that higher frequency modes are only weakly excited, if at all, in general. However, our results indicate that when multiple localized modes, and sometimes multiple delocalized ones, are present, they can be significantly excited, suppressing any resonant behavior. Deriving equations that describe how energy is redistributed among all modes at each bounce would advance Campbell's original description to a more comprehensive framework. Progress in this direction has been made through the study of wobbling kink collisions \cite{PhysRevD.103.045003} and the inclusion of several degrees of freedom in the collective coordinate formalism of kink interactions \cite{Blaschke, Navarro2023Inclusion, Adam2023Relat, Jiang}.

\section*{Acknowledgements}

JGFC acknowledges financial support from Fundação de Amparo a Ciência e Tecnologia do Estado de Pernambuco (FACEPE), grant no. BFP-0013-1.05/23. AM acknowledges financial support from Conselho Nacional de Desenvolvimento Científico e Tecnológico (CNPq), Grant no. 309368/2020-0. CESS and AM also acknowledge financial support from Coordenação de Aperfeiçoamento de Pessoal de Nível Superior (CAPES). Part of the simulations presented here were performed in the supercomputer SDumont at the Brazilian agency LNCC (Laboratório Nacional de Computação Científica).

\appendix

\section{Potential and kink parameters}
\label{ap:rel}

Imposing the continuity condition on the potential yields to the following four relations
\begin{align}
A^2 &= B^2 \frac{(\phi_0 - \phi_1)}{\phi_1},\\
V_+ &= \frac{1}{2} B^2 [\phi_1 (\phi_0 - \phi_1) + (\phi_0 - \phi_1)^2],\\
\lambda &= \frac{2 V_+}{B^2(\phi_2 - \phi_0)} + \phi_0,\\
C^2 &= B^2 \frac{(\phi_0 - \phi_2)}{(\phi_2 - \lambda)}.
\end{align}

Imposing the same condition on the kink profile leads to the following relations
\begin{align}
K &= \frac{\sqrt{2 V_+}}{B},\\
\theta_0 &= \sin^{-1} \left[ \frac{1}{K} (\phi_1 - \phi_0) \right],\\
x_1 &= \frac{1}{A} \ln{\phi_1},\\
x_2 &= x_1 + \frac{1}{B} \sin^{-1}\left(\frac{\phi_2 - \phi_0}{K} \right) - \frac{\theta_0}{B}.
\end{align}

\section{Solutions of one and two asymmetric square-wells}
\label{ap:modes}

In this appendix, we introduce the momenta as $k_1^2 = A^2 - \omega^2$, $k_2^2 = B^2 + \omega^2$, and $k_3^2 = C^2 - \omega^2$. Moreover, the square-well size is defined as $L = x_2 - x_1$. Once the vibrational mode frequency is obtained from the corresponding transcendental equation, we may express the asymmetric square-well localized modes as
\begin{equation}
\eta(x) =
\begin{cases}
G e^{k_1(x- x_1)}, &x < x_1, \\
H_1 \sin [k_2(x - x_1)] + H_2 \cos [k_2(x - x_1)], &x_1 < x < x_2,\\
I e^{-k_3(x-x_2)}, &x > x_2.
\end{cases}
\end{equation}
The constants are obtained by requiring continuity of $\eta$ and its first derivative. They are
\begin{align}
H_1 &= \frac{k_1}{k_2} G, \\
H_2 &= G, \\
I &= G \left\{ \frac{k_1}{k_2} \sin [k_2(x_2 - x_1)] + \cos[k_2 (x_2 - x_1)] \right\}.
\end{align}
Here and in the following expressions, the parameters $G$ can be fixed by requiring normalization of the modes.

Now, we make the following redefinition $k_1^2=\omega^2-A^2$. Then, the even delocalized are expressed as
\begin{equation}
\eta(x) =
\begin{cases}
G \exp\left[k_3\left(x+\frac{D}{2}+L\right)\right], &x < -\frac{D}{2}-L, \\
H_1 \sin \left[k_2\left(x +\frac{D}{2}\right)\right] + H_2 \cos \left[k_2\left(x +\frac{D}{2}\right)\right], &-\frac{D}{2}-L < x < -\frac{D}{2},\\
I \cos (k_1 x), &-\frac{D}{2} < x < \frac{D}{2},\\
-H_1 \sin \left[k_2\left(x - \frac{D}{2}\right)\right] + H_2 \cos \left[k_2\left(x - \frac{D}{2}\right)\right], &\frac{D}{2} < x < \frac{D}{2}+L,\\
G \exp\left[-k_3\left(x-\frac{D}{2}-L\right)\right], &x > \frac{D}{2}+L.
\end{cases}
\end{equation}
Requiring continuity of $\eta$ and its first derivative, we obtain
\begin{align}
H_1 &= G\left[\frac{k_3}{k_2}\cos(k_2L)-\sin(k_2L)\right], \\
H_2 &= G\left[\cos(k_2L)+\frac{k_3}{k_2}\sin(k_2L)\right], \\
I &= \frac{G}{\cos(k_1D/2)}\left[\cos(k_2L)+\frac{k_3}{k_2}\sin(k_2L)\right].
\end{align}
Likewise, the odd delocalized are expressed as
\begin{equation}
\eta(x) =
\begin{cases}
G \exp\left[k_3\left(x+\frac{D}{2}+L\right)\right], &x < -\frac{D}{2}-L, \\
H_1 \sin \left[k_2\left(x +\frac{D}{2}\right)\right] + H_2 \cos \left[k_2\left(x +\frac{D}{2}\right)\right], &-\frac{D}{2}-L < x < -\frac{D}{2},\\
I \sin (k_1 x), &-\frac{D}{2} < x < \frac{D}{2},\\
H_1 \sin \left[k_2\left(x - \frac{D}{2}\right)\right] - H_2 \cos \left[k_2\left(x - \frac{D}{2}\right)\right], &\frac{D}{2} < x < \frac{D}{2}+L,\\
-G \exp\left[-k_3\left(x-\frac{D}{2}-L\right)\right], &x > \frac{D}{2}+L.
\end{cases}
\end{equation}
Requiring continuity of $\eta$ and its first derivative, we obtain the same expression for $H_1$ and $H_2$, while the remaining constant is
\begin{equation}
I = -\frac{G}{\sin(k_1D/2))}\left[\cos(k_2L)+\frac{k_3}{k_2}\sin(k_2L)\right].
\end{equation}

The half-localized (or half-scattering) modes for an isolated kink obeying $A<\omega<C$ are as follows
\begin{equation}
\eta(x)=
\begin{cases}
Ge^{ik_1(x-x_1)} + G^* e^{-ik_1(x-x_1)}, \ \ x < x_1, \\
H_1 \sin{[k_2(x-x1)]} + H_2 \cos{[k_2(x-x1)]}, \ \ x_1 < x < x_2 ,\\
I e^{-k_3(x-x_1)}, \ \ x > x_2,
\end{cases}
\end{equation}
where
\begin{align}
H_1 &= \left[-\sin{(k_2 L)} + \dfrac{k_3}{k_2} \cos{(k_2 L)} \right] I, \\
H_2 &= \left[-\cos{(k_2 L)} - \dfrac{k_3}{k_2} \sin{(k_2 L)} \right] I,\\
G &= \dfrac{1}{2} \left( H_2 - i \dfrac{k_2}{k_1}H_1 \right) ,\\
G^* &= \dfrac{1}{2} \left( H_2 + i \dfrac{k_2}{k_1}H_1 \right) .
\end{align}

\section{Numerical Method}
\label{ap:num}

The partial differential field equation was solved utilizing a second-order finite differences scheme. Such methods are expected to handle discontinuities better than higher-order ones. The field is discretized as $\phi_{i,j}\equiv\phi(x=-L_{\text{box}}+i\delta x,t=j\delta t)$ with box size $2L_{\text{box}}=600.0$ and periodic boundary conditions. In all simulations, we have utilized $\delta x = 10^{-2}$ and $\delta t = 10^{-3}$, leading to indices $i=0,1,...,6\times 10^4$, $j=0,1,...,4\times 10^5$. Then, the partial derivatives are computed as
	\begin{equation}
	\frac{\partial^2\phi_{i,j}}{\partial t^2}  \approx \frac{\phi_{i,j+1} - 2 \phi_{i,j} + \phi_{i,j-1}}{(\delta t)^2},
	\end{equation}
	\begin{equation}
	\frac{\partial^2\phi_{i,j}}{\partial x^2} \approx \frac{\phi_{i+1,j} - 2 \phi_{i,j} + \phi_{i-1,j}}{(\delta x)^2}.
	\end{equation}
Additionally, damping proportional to a bump function was added to ensure that no radiation returns to the system. This function remains zero everywhere except at the boundaries, where it has a smooth bump shape. Our approach included conserving energy as a method to gauge the overall simulation accuracy, with all presented results exhibiting a maximum relative error of order $10^{-4}$.

\end{document}